\documentclass[aps,floatfix,amsmath,nofootinbib,amssymb,superscriptaddress]{revtex4}

\usepackage{booktabs} 

\usepackage{overpic}
\usepackage{amssymb}
\usepackage{indentfirst}
\usepackage{feynmf}   
\usepackage{slashed}  
\usepackage{cases}
\usepackage{color}
\usepackage{multirow}
\usepackage{epstopdf}
\usepackage{graphicx,color,bm}
\usepackage{epstopdf}

\usepackage[colorlinks,
            citecolor=green,
            anchorcolor=red,
            menucolor=red,
            linkcolor=red,
            filecolor=red,
            runcolor=red,
            urlcolor=blue,
            frenchlinks=red]{hyperref}

\begin{document}

\title{Gluon gravitational form factors of the proton from near-threshold charmonium photoproduction}

\author{Chentao Tan}\affiliation{School of Physics, Southeast University, Nanjing
211189, China}

\author{Zhun Lu}
\email{zhunlu@seu.edu.cn}
\affiliation{School of Physics, Southeast University, Nanjing 211189, China}

\begin{abstract}
	
Motivated by near-threshold $J/\psi$ photoproduction measurements, we investigate the gluon contributions to the gravitational form factors (GFFs) of the proton associated with the energy-momentum tensor in quantum chromodynamics. We calculate the gluon GFFs and the corresponding mechanical properties of the proton within a light-front spectator model, in which the proton is described as an active gluon accompanied by a spectator particle. The model parameters are determined from a global fit to the differential cross section for $J/\psi$ photoproduction. The resulting gluon GFFs are in reasonable agreement with recent lattice QCD determinations. We further calculate the total cross section and find reasonable agreement with the available experimental data. Using the extracted gluon GFFs, we investigate the two-dimensional light-front gluon distributions inside the proton, such as the mass, angular momentum, pressure, and shear force distributions. We also study the gluonic contributions to the two-dimensional Galilean energy density, radial and tangential pressures, isotropic pressure, and pressure anisotropy.

\end{abstract}

\maketitle

\section{Introduction}

Understanding the proton structure in terms of its quark and gluon degrees of freedom remains a central challenge in particle physics~\cite{Aschenauer:2014twa,AbdulKhalek:2021gbh,Anderle:2021wcy,
Burkert:2018nvj,Accardi:2023chb}. 
Form factors provide essential information about the internal structure of hadrons. 
While the weak and electromagnetic form factors of the proton are well established, its gravitational form factors (GFFs)~\cite{Polyakov:2018zvc,Pagels:1966zza,Kobzarev:1962wt,Lorce:2018egm,Burkert:2018bqq,Shanahan:2018nnv} remain much less constrained. 
Particular attention has been devoted to the mechanical properties of the proton, including the distributions of mass, angular momentum, pressure, and shear force carried by quarks and gluons~\cite{Lorce:2018egm,Polyakov:2002yz,Polyakov:2018zvc,Burkert:2023wzr,Goharipour:2025yxm,Kharzeev:2021qkd,Ji:1994av,Ji:1995sv,Lorce:2017xzd,Rodini:2020pis,Metz:2020vxd,EuropeanMuon:1987isl,Ji:1996ek,Leader:2013jra,Lorce:2021xku}. 
These quantities are encoded in GFFs, which are defined through the parametrization of matrix elements of the QCD energy-momentum tensor (EMT) in proton states~\cite{Pagels:1966zza,Kobzarev:1962wt,Polyakov:2018zvc}.

Generalized parton distributions (GPDs)~\cite{Muller:1994ses,Ji:1996ek,Goeke:2001tz,Diehl:2003ny,
Belitsky:2005qn,Radyushkin:1997ki} provide a unified framework for describing hadronic form factors. 
In particular, the GFFs associated with the symmetric, traceless part of the EMT are given by second Mellin moments of GPDs. GPDs can be accessed in high-energy exclusive processes such as deeply virtual Compton scattering (DVCS)~\cite{dHose:2016mda,Kumericki:2016ehc,Ji:1996nm,Radyushkin:1996nd} and hard exclusive meson production~\cite{Collins:1996fb,Favart:2015umi,Radyushkin:1996ru}. 
The quark GFFs for the proton have been extracted at Jefferson Lab (JLab) from DVCS~\cite{CLAS:2007clm,Burkert:2018bqq,Pasquini:2014vua,CLAS:2015uuo}, while further constraints can be obtained from GPD measurements at future facilities, including the JLab 12 GeV program~\cite{Dudek:2012vr,Burkert:2018nvj,Arrington:2021alx} and Electron-Ion Colliders (EICs)~\cite{Aschenauer:2014twa,Accardi:2012qut,AbdulKhalek:2021gbh,
Anderle:2021wcy,Joosten:2018gyo,Abir:2023fpo,Gryniuk:2020mlh,Wang:2023thy}.

Substantial progress has also been made on the theoretical evaluation of the quark GFFs of the proton. These efforts include analytical approaches such as the multipole model~\cite{Lorce:2018egm} and chiral perturbation theory~\cite{Chen:2001pva,Belitsky:2002jp,Dorati:2007bk}, as well as first-principles lattice QCD simulations~\cite{Hagler:2003jd,Gockeler:2003jfa,LHPC:2010jcs,LHPC:2007blg,
QCDSF-UKQCD:2007gdl,Deka:2013zha}. 
Complementary phenomenological extractions and model calculations~\cite{Schweitzer:2002nm,Wakamatsu:2006dy,
Goeke:2007fq,Goeke:2007fp,Wakamatsu:2007uc,Neubelt:2019sou,Cebulla:2007ei,
Kim:2012ts,Polyakov:2018exb,Chakrabarti:2015lba,Chakrabarti:2020kdc,
Chakrabarti:2021mfd,
Chakrabarti:2016mwn,Kumar:2017dbf,Mondal:2015fok,Choudhary:2022den,
Mondal:2016xsm,Abidin:2009hr,Fujita:2022jus,Nair:2024fit,Anikin:2019kwi,
Dehghan:2025ncw} have been widely pursued. Perturbative QCD computations at large momentum transfer~\cite{Tong:2021ctu,Tong:2022zax} and global phenomenological analyses~\cite{Pasquini:2014vua,Cao:2023ohj,Hatta:2018sqd,Tanaka:2018nae,
Goharipour:2025lep,Cao:2024zlf,Wang:2023fmx} further enrich the theoretical landscape of quark GFFs.

In contrast, the gluon contributions to the proton GFFs remain much less constrained. 
Processes commonly used to access quark GPDs have limited sensitivity to gluons. 
Near-threshold exclusive electroproduction or photoproduction of heavy quarkonia, such as $J/\psi$ and $\Upsilon$, has therefore emerged as a promising channel for probing gluon GFFs at finite momentum transfer~\cite{Wang:2022ndz,Guo:2023pqw,Guo:2023qgu,Hatta:2019lxo,
Boussarie:2020vmu,Pentchev:2025qyn,Sun:2021gmi,Hatta:2018ina,Guo:2021ibg}. 
In the heavy-quark limit, light-quark contributions are strongly suppressed, and the process is dominated by two-gluon exchange~\cite{Wang:2022vhr,Brodsky:1997gh,Sun:2021gmi}. 
Recent near-threshold measurements of $\gamma p \rightarrow J/\psi\, p$ by the GlueX~\cite{GlueX:2019mkq,GlueX:2023pev} and $J/\psi$-007~\cite{Duran:2022xag} collaborations at JLab have stimulated substantial theoretical interest.
 
Future measurements at JLab and EICs are expected to provide increasingly precise $J/\psi$ and $\Upsilon$ data~\cite{AbdulKhalek:2021gbh,Anderle:2021wcy}. 
Theoretical studies of gluon GFFs remain less extensive than those of quark GFFs, partly because most phenomenological nucleon models do not include explicit gluonic degrees of freedom. 
Nevertheless, gluon GFFs have been investigated using lattice QCD~\cite{Shanahan:2018pib,Pefkou:2021fni,Hackett:2023rif,Shanahan:2018nnv}, simple multipole models~\cite{Guo:2023pqw,Wang:2022ndz,Guo:2021ibg}, holographic QCD~\cite{Mamo:2019mka,Mamo:2022eui,Mamo:2021krl}, an extended holographic light-front QCD framework~\cite{deTeramond:2021lxc,Gurjar:2022jkx}, a light-front spectator model motivated by soft-wall AdS/QCD~\cite{Sain:2025kup}, the dressed quark model~\cite{More:2023pcy}, and the Bethe-Salpeter equation approach~\cite{Yao:2024ixu}. 

The GFFs encode spatial densities through Fourier transforms~\cite{Polyakov:2018zvc,Lorce:2018egm}. 
However, the physical meaning of these densities must be carefully distinguished. 
Specifically, the Breit frame densities~\cite{Polyakov:2018zvc,Lorce:2018egm}, obtained via the three-dimensional (3D) Fourier transform at fixed time (instant form), are understood as quasi-probabilistic densities~\cite{Lorce:2017wkb,Lorce:2020onh,Lorce:2018zpf,Lorce:2021gxs}, since a hadron cannot be spatially localized~\cite{Miller:2018ybm,Jaffe:2020ebz}. 
In contrast, the 2D Fourier transform at fixed light-front time provides a correct alternative~\cite{Xu:2024hfx,Miller:2007uy,Burkardt:2002hr,Miller:2009sg,
Carlson:2007xd,Miller:2018ybm,Freese:2021czn,Kim:2021jjf,Mondal:2017lph,
Lorce:2018egm}.
The resulting 2D distributions acquire a genuine probabilistic interpretation, with the relativistic corrections being suppressed kinematically, and the large-mass assumption inherent in the Breit frame being eliminated~\cite{Epelbaum:2022fjc}. 
Moreover, the densities on the light-front (or in the infinite momentum frame) are not affected by Lorentz contraction due to the on-shell condition of the incoming and outgoing nucleon states, and remain fully relativistic due to the Galilean symmetry inherent in the light-front framework.
One-to-one relations between light-front and Breit frame distributions can be established via the Abel transformation~\cite{Panteleeva:2021iip}, and definitions of densities in other frames were also reported in Ref.~\cite{Lorce:2018egm}. 

GFFs also encode spatial distributions through Fourier transforms~\cite{Polyakov:2018zvc,Lorce:2018egm}. 
The physical interpretation of these distributions, however, depends on the reference frame. 
In the Breit frame, three-dimensional (3D) Fourier transforms at fixed time (instant form) yield quasi-probabilistic densities~\cite{Polyakov:2018zvc,Lorce:2018egm,Lorce:2017wkb,Lorce:2020onh,
Lorce:2018zpf,Lorce:2021gxs}, because a hadron cannot be perfectly localized in space~\cite{Miller:2018ybm,Jaffe:2020ebz}. 
By contrast, two-dimensional (2D) Fourier transforms at fixed light-front time provide an alternative description~\cite{Xu:2024hfx,Miller:2007uy,Burkardt:2002hr,Miller:2009sg,Carlson:2007xd,Miller:2018ybm,Freese:2021czn,Kim:2021jjf,Mondal:2017lph,Lorce:2018egm}. 
The resulting 2D distributions admit a probabilistic interpretation, with relativistic corrections kinematically suppressed and the large-mass assumption associated with the Breit-frame description avoided~\cite{Epelbaum:2022fjc}. 
Moreover, light-front distributions are not subject to Lorentz contraction and retain a fully relativistic character owing to the Galilean symmetry of the light-front framework. 
One-to-one relations between light-front and Breit-frame distributions can be established through an Abel transformation~\cite{Panteleeva:2021iip}, while definitions of distributions in other frames have also been discussed in Ref.~\cite{Lorce:2018egm}.

The GFFs and related distributions can explain the mechanical stability of hadrons. 
In this paper, we employ a light-front spectator model for the proton~\cite{Lu:2016vqu} to calculate the gluon GFFs, where the proton is regarded as a two-body system composed of an active gluon and a spectator particle that carries the quantum numbers of the three valence quarks. 
Model parameters are determined by a global fit to the differential cross section of near-threshold $J/\psi$ photoproduction~\cite{GlueX:2019mkq,GlueX:2023pev,Duran:2022xag}. 
The model predictions for the gluon GFFs and the total cross section are consistent with recent lattice QCD determinations~\cite{Pefkou:2021fni,Hackett:2023rif} and experimental measurements~\cite{GlueX:2019mkq,GlueX:2023pev}. 
Using these GFFs, we study the 2D light-front gluon densities including the mass, angular momentum, pressure, shear force, and radial and tangential forces, as well as the 2D Galilean energy density, radial and tangential pressures, isotropic pressure, and pressure anisotropy.

The remainder of this paper is organized as follows. 
In Sec.~\ref{Sec:2}, we introduce the gluon GFFs through the gluon sector of the QCD EMT and give the expressions for the differential and total cross sections for near-threshold heavy-quarkonium photoproduction. 
In Sec.~\ref{Sec:3}, we calculate the gluon GFFs in the light-front spectator model. 
In Sec.~\ref{Sec:4}, we determine the model parameters from a global fit to the differential cross section for $J/\psi$ photoproduction, compare the resulting GFFs and total cross section with lattice QCD and experimental results, and investigate the 2D light-front and 2D Galilean gluon distributions in the proton. 
A summary is given in Sec.~\ref{Sec:5}.

\section{The near-threshold heavy quarkonium photoproduction and gluon GFFs}\label{Sec:2}

We begin with a brief review of the GPD framework for the near-threshold heavy quarkonium photoproduction~\cite{Guo:2021ibg,Guo:2023pqw,Guo:2023qgu}. 
The kinematics are described by the Mandelstam invariants
\begin{align}
	s = (p+p_\gamma)^2 \geq (M+M_V)^2, \qquad 
	t = (p^\prime-p)^2 = (p_V-p_\gamma)^2,
\end{align}
where $s = W^2$ is the squared center-of-mass energy, $t=q^2$ is the squared four-momentum transfer to the proton, $M$ is the proton mass, and $M_V$ is the produced meson mass. 
The four-momenta of the incoming photon, incoming proton, outgoing proton, and outgoing meson are denoted by $p_\gamma$, $p$, $p^\prime$, and $p_V$, respectively, satisfying the on-shell conditions: 
\begin{align}
	p^{\prime 2}=p^2=M^2,\qquad p_\gamma^2=0,\qquad p_V^2=M_V^2,
\end{align}	 
where $p_\gamma^2=0$ corresponds to photoproduction. 
In Ref.~\cite{Guo:2021ibg}, this process was formulated in the photon-proton center-of-mass frame; the resulting expressions are Lorentz invariant. 
In the heavy-quark limit $M_V \gg M$ (which implies $W\gg M$), the incoming proton moves nearly along the light-front $+$ direction, defined by $x^{\pm} \equiv x^0 \pm x^z$. 
At threshold, $\sqrt{s} \to M+M_V$, the squared momentum transfer becomes
\begin{align}
	-t_{\text{th}}=\frac{MM_V^2}{M+M_V},
\end{align}	
which is of order $MM_V$ in the heavy-quark limit.

A key feature of near-threshold kinematics is that $|t|$ is large, and the skewness parameter $\xi$, defined as $\xi= -q\cdot p_\gamma/(2P\cdot p_\gamma) = (p^+-p^{\prime+})/(p^++p^{\prime+})$ with the average proton four-momentum $P=(p^\prime+p)/2$, approaches 1 in the heavy-quark limit. 
Combined with QCD factorization in the GPD framework, this ensures that the photoproduction amplitude of heavy quarkonium is dominated by the second Mellin moments of gluon GPDs. 
After averaging over the initial proton spin and summing over the final proton spin and the photon and meson polarizations, the $t$-dependent differential cross section and total cross section can be written as~\cite{Guo:2021ibg,Guo:2023pqw,Guo:2023qgu}:
\begin{align} 
	\frac{d\sigma}{dt}&=\frac{\alpha_\text{EM}e^2_\text{Q}}{4(W^2-M^2)^2}
	\frac{(16\pi\alpha_S)^2}{3M_V^3}|\psi_\text{NR}(0)|^2|G(t,\xi)|^2,
	\label{eq:dsigma}\\
	\sigma&=\int^{t_{\text{max}}}_{t_{\text{min}}} dt \left(\frac{d\sigma}{dt}\right),
	\label{eq:sigma}
\end{align}	
where $\psi_{\text{NR}}(0)$ stands for the non-relativistic quarkonium wave function at the origin in the non-relativistic QCD framework~\cite{Bodwin:1994jh}, which is associated with the quarkonium distribution amplitude, and the upper and lower limits ($t_{\text{max}}$ and $t_{\text{min}}$) of the integral correspond to the kinematically allowed $t$-range on the $(W,-t)$ plane for heavy quarkonium production. 

In the unpolarized case, the squared hadronic matrix element $|G(t,\xi)|^2$ reads
\begin{align}
	|G(t,\xi)|^2=\frac{4}{\xi^4}\left\{\left(1-\frac{t}{4M^2}\right)
	E_2^2-2E_2(H_2+E_2)+(1-\xi^2)(H_2+E_2)^2\right\},
	\label{eq:Gt}
\end{align}	
where $H_2\equiv H_2(t,\xi)$ and $E_2\equiv E_2(t,\xi)$ are defined as the second moments of GPDs,
\begin{align}
	\int_0^1 dx x H_g(x,\xi,t)&=A_g(t)+\xi^2 D_g(t)\equiv H_2(t,\xi),\nonumber\\
	\int_0^1 dx x E_g(x,\xi,t)&=B_g(t)-\xi^2 D_g(t)\equiv E_2(t,\xi),
	\label{eq:H2E2}
\end{align}	
with $A_g(t)$, $B_g(t)$, and $D_g(t)$ denoting the gluon GFFs, and $x$ the longitudinal momentum fraction of the active gluon. 
The renormalization scales of the wave function and GPDs are omitted and are understood to be of order $M_V$.

For a spin-$1/2$ target, the off-forward nucleon matrix element of the quark and gluon parts of a local, gauge invariant, and symmetric QCD EMT $T^{\mu\nu}(0)$ can be parameterized in terms of the following four GFFs~\cite{Ji:2012vj,Harindranath:2013goa}:
\begin{align}
	\langle p^\prime,\lambda^\prime|T_i^{\mu\nu}(0)|p,\lambda\rangle=&\bar{u}^\prime(p^\prime,\lambda^\prime)\bigg[-B_i(q^2)\frac{P^\mu P^\nu}{M}+(A_i(q^2)+B_i(q^2))\frac{\gamma^\mu P^\nu+\gamma^\nu P^\mu}{2}+D_i(q^2)\frac{q^\mu q^\nu-q^2 g^{\mu\nu}}{4M}\nonumber\\
	&+\bar{C}_i(q^2)Mg^{\mu\nu}\bigg]u(p,\lambda),
\end{align}
where the subscript $i$ represents the quark or gluon, a generic four-momentum $a^\mu$ is represented by its light-front components $[a^+,a^-,\bm{a}_\perp]$, the nucleon states and spinors are normalized by $\langle p^\prime,\lambda^\prime|p,\lambda\rangle=2p^0(2\pi)^3\delta_{\lambda^\prime\lambda}\delta^{(3)}(\bm{p}^\prime-\bm{p})$ and $\bar{u}(p^\prime,\lambda^\prime)u(p,\lambda)=2M\delta_{\lambda^\prime\lambda}$, and $D_i(q^2)=4C_i(q^2)$. 
The total EMT is then simply obtained by summing over all the constituent types: $T^{\mu\nu}=\sum_{i=q,g} T_i^{\mu\nu}$. 
In general, the $A$ form factor characterizes the momentum distribution of quarks or gluons inside the hadron, whereas the $D$ form factor encodes its mechanical properties.
We work in the symmetric Drell-Yan frame with vanishing average transverse momentum, $\bm{P}_\perp=0$, and $q^+=0$. Thus, $q=(0,0,\bm{q}_\perp)$ and $-t=-q^2=\bm{q}_\perp^2$.

The gauge invariant symmetric QCD EMT is expressed as~\cite{Harindranath:1997kk}
\begin{align}
	T^{\mu\nu}=\frac{1}{2}\bar{\psi}i[\gamma^\mu D^\nu+\gamma^\nu D^\mu]\psi-F^{\mu\lambda a}F^{\nu}_{~\lambda a}+\frac{1}{4}g^{\mu\nu}(F_{\lambda\sigma a})^2-g^{\mu\nu}\bar{\psi}(i\gamma^\lambda D_\lambda-m)\psi,\label{eq:Tmunu}
\end{align}	
where the covariant derivative is defined as $iD^\mu=i\overleftrightarrow{\partial}^\mu+gA^\mu$ with $\alpha(i\overleftrightarrow{\partial}^\mu)\beta=\frac{i}{2}\alpha(\partial^\mu\beta)-\frac{i}{2}(\partial^\mu\alpha)\beta$, and $F_a^{\mu\nu}$ is the field strength tensor
\begin{align}
	F_a^{\mu\nu}=\partial^\mu A_a^\nu-\partial^\nu A^\mu_a +gf^{abc}A_b^\mu A_c^\nu,
\end{align}	

In this work, we focus on the gluon sector of the EMT in Eq.~(\ref{eq:Tmunu})
\begin{align}
	T_g^{\mu\nu}=-F^{\mu\lambda a}F^{\nu}_{~\lambda a}+\frac{1}{4}g^{\mu\nu}(F_{\lambda\sigma a})^2.	
\end{align}	
In order to calculate the gluon GFFs, the matrix element of the EMT is defined as
\begin{align}
	\mathcal{M}_{\lambda\lambda^\prime}^{\mu\nu}=\frac{1}{2}[\langle p^\prime,\lambda^\prime|T^{\mu\nu}_g(0)|p,\lambda\rangle].\label{eq:Mmunu}
\end{align}	
The GFFs $A_g(q^2)$ and $B_g(q^2)$ can be extracted from the $T^{++}_g$ component of the EMT, while the GFFs $D_g(q^2)$ and $\bar{C}_g(q^2)$ can be extracted from the $T_g^{\mu1}$ component and the transverse component $T_g^{jj}$, with $j=1,2$. 
Four GFFs can then be obtained using Eq.~(\ref{eq:Mmunu}) as follows~\cite{More:2023pcy,Sain:2025kup}:
\begin{align}	\mathcal{M}_{\uparrow\uparrow}^{++}+\mathcal{M}_{\downarrow\downarrow}^{++}
&=2(p^+)^2A_g(q^2),\label{eq:M1}\\	\mathcal{M}_{\uparrow\downarrow}^{++}+\mathcal{M}_{\downarrow\uparrow}^{++}
&=\frac{iq_\perp^{(2)}}{M}(p^+)^2B_g(q^2),\label{eq:M2}\\
	q_\mu \mathcal{M}_{\uparrow\downarrow}^{\mu 1} + q_\mu \mathcal{M}_{\downarrow\uparrow}^{\mu 1}
	&=-iq_\perp^{(1)}q_\perp^{(2)}M\bar{C}_g(q^2),\label{eq:M3}\\	\mathcal{M}_{\uparrow\downarrow}^{11}+\mathcal{M}_{\uparrow\downarrow}^{22}
+\mathcal{M}_{\downarrow\uparrow}^{11}+\mathcal{M}_{\downarrow\uparrow}^{22}
&=i\left[B_g(q^2)\frac{(\bm{q}_\perp)^2}{4M}-D_g(q^2)\frac{3(\bm{q}_\perp)^2}{4M}
	+\bar{C}_g(q^2)2M\right]q_\perp^{(2)}\label{eq:M4},
\end{align}	
where the GFF $A_g(q^2)$ involves the sum of the helicity-conserving matrix elements, while the other three GFFs involve the sum of the helicity-flip matrix elements, with proton helicities $(\lambda,\lambda^\prime)\equiv\{\uparrow,\downarrow\}=\{+\frac{1}{2},-\frac{1}{2}\}$. 
Furthermore, the GFFs $A_g(q^2)$ and $B_g(q^2)$ are analogous to the Dirac and Pauli form factors, respectively.

\section{Gluon GFFs in the light-front spectator model}\label{Sec:3}

In this section, we calculate the gluon GFFs of the proton using the overlap representation of the matrix elements of EMT components within a light-front spectator model~\cite{Lu:2016vqu}. 
The minimum Fock state for a proton target containing a gluon is $|qqqg\rangle$. 
As the four-body system is rather complicated, here we resort to a more phenomenological approach, assuming that the three valence quarks can be grouped into a spectator particle~\cite{Lu:2016vqu,Bacchetta:2020vty}. 
Then the proton can be viewed as a two-body system composed of an active gluon $g$ and a spectator particle $X$:
\begin{align}
	|p,S \rangle \rightarrow |g_{s_g}X_{s_X}(uud)\rangle ,
\end{align}
where $s_g$ and $s_X$ represent the spins of the gluon and the spectator particle, respectively. 
In this model, the spectator particle can be seen as a nucleon in the naive quark model with spin $s_X=1/2$, or it can be a state with $s_X=3/2$. 
If the orbital angular momentum among the quarks in the spectator three-quark system is considered, the spin of the spectator can be higher than $3/2$. 
In principle, the projections of the spectator spin in the $z$-direction can be $s_{\rm{X}}^z=\pm1/2$, $\pm3/2$, and so on. 
Since the spectator here is an isolated and thermodynamically balanced quantum system, its probability of being in the ground state is always greater than that in an excited state. 
Therefore, we only consider the $s_X^z=\pm1/2$ components and ignore the contributions from higher-spin spectators, following the convention in Refs.~\cite{Lu:2016vqu,Bacchetta:2020vty}.

For a proton with helicities $\lambda=\pm1/2$, the Fock-state expansions are expressed as
\begin{align}
	|\Psi^{\uparrow(\downarrow)}_{\text{two\, particle}}(p^+,\bm{p}_\perp=\bm{0}_\perp) \rangle =&\int \frac{d^2\bm{k}_\perp dx}{16\pi^3\sqrt{x(1-x)}}\nonumber \\
	&\times \left[\psi^{\uparrow(\downarrow)}_{+1+\frac{1}{2}}(x,\bm{k}_\perp)
	\left|+1,+\frac{1}{2};xp^+,\bm{k}_\perp \right\rangle+\psi^{\uparrow(\downarrow)}_{+1-\frac{1}{2}}(x,\bm{k}_\perp)
	\left|+1,-\frac{1}{2};xp^+,\bm{k}_\perp \right\rangle \nonumber \right.\\		&\left.+\psi^{\uparrow(\downarrow)}_{-1+\frac{1}{2}}(x,\bm{k}_\perp)
	\left|-1,+\frac{1}{2};xp^+,\bm{k}_\perp \right\rangle+\psi^{\uparrow(\downarrow)}_{-1-\frac{1}{2}}(x,\bm{k}_\perp)
	\left|-1,-\frac{1}{2};xp^+,\bm{k}_\perp \right\rangle \right],
\end{align}
where $\bm{k}_\perp$ is the transverse momentum carried by the active gluon, and $\psi_{\lambda_g\lambda_X}^{\lambda}(x,\bm{k}_\perp)$ are the light-front wave functions (LFWFs) corresponding to the two-particle states $|\lambda_g,\lambda_X;xp^+,\bm{k}_\perp\rangle$, with $\lambda_g=\pm1$ and $\lambda_X=\pm1/2$ being the helicities of the gluon and the spectator, respectively. 
These LFWFs have the following forms~\cite{Lu:2016vqu}:
\begin{align} \psi^\uparrow_{+1+\frac{1}{2}}(x,\bm{k}_\perp)&=
-\sqrt{2}\frac{-k^1_\perp+ik^2_\perp}{x(1-x)}\phi,\nonumber\\ \psi^\uparrow_{+1-\frac{1}{2}}(x,\bm{k}_\perp)&=
-\sqrt{2}\left(M-\frac{M_X}{1-x}\right)\phi,\nonumber\\ \psi^\uparrow_{-1+\frac{1}{2}}(x,\bm{k}_\perp)&=
-\sqrt{2}\frac{+k^1_\perp+ik^2_\perp}{x}\phi,\nonumber\\
	\psi^\uparrow_{-1-\frac{1}{2}}(x,\bm{k}_\perp)&=0,
	\label{eq:wavefunction+}
\end{align}
for the proton with $\lambda=+1/2$, and
\begin{align}
	\psi^\downarrow_{+1+\frac{1}{2}}(x,\bm{k}_\perp)&=0,\nonumber\\ \psi^\downarrow_{+1-\frac{1}{2}}(x,\bm{k}_\perp)&=
-\sqrt{2}\frac{-k^1_\perp+ik^2_\perp}{x}\phi,\nonumber\\	\psi^\downarrow_{-1+\frac{1}{2}}(x,\bm{k}_\perp)&=
-\sqrt{2}\left(M-\frac{M_X}{1-x}\right)\phi,\nonumber\\	\psi^\downarrow_{-1-\frac{1}{2}}(x,\bm{k}_\perp)&=
-\sqrt{2}\frac{+k^1_\perp+ik^2_\perp}{x(1-x)}\phi,
	\label{eq:wavefunction-}
\end{align}
for the proton with $\lambda=-1/2$, where
\begin{align}
	\phi(x,\bm{k}_\perp)=\frac{\lambda \sqrt{x} (1-x)}{x(1-x)M^2-(1-x)\bm{k}^2_\perp-x(\bm{k}_\perp^2+M_X^2)}
\end{align}
represents the wave function in momentum space, $M_X$ is the spectator mass, and the gluon mass $M_g$ is taken as $M_g=0$. 
The wave functions in Eqs.~(\ref{eq:wavefunction+}-\ref{eq:wavefunction-}) are inspired by the wave functions of electron Fock states~\cite{Brodsky:2000ii}, whose Fock state consists of a spin-1 photon and a spin-1/2 electron. 
In order to simulate the nonperturbative effect of the nucleon-gluon-spectator vertex, we choose the Brodsky-Hwang-Lepage prescription for the coupling $\lambda$ of the vertex~\cite{Brodsky:1982nx}:
\begin{align}
	\lambda \rightarrow N_\lambda \exp(-\frac{\mathcal{M}^2}{2\beta^2_1}),
\end{align}
where $N_\lambda$ is a strength parameter for the vertex, $\beta_1$ is a cutoff parameter, and $\mathcal{M}$ represents the invariant mass of the two-particle system
\begin{align}
	\mathcal{M}^2=\frac{\bm{k}^2_\perp}{x}+\frac{\bm{k}^2_\perp+M_X^2}{1-x}.
\end{align}
 
The derivations of the gluon GFFs $A_g(q^2)$ and $B_g(q^2)$ in Eqs.~(\ref{eq:M1}-\ref{eq:M2}) require the nucleon matrix element of the local operator $T_g^{++}$, which has an exact overlap representation in terms of the LFWFs of the nucleon~\cite{Sain:2025kup}:
\begin{align}
	\langle p^\prime,\lambda^\prime | T_g^{++} | p,\lambda \rangle
	= 2\left(p^+\right)^2
	\int \frac{\mathrm{d}^2\bm{k}_\perp \mathrm{d}x}{16\pi^3}
	x \sum_{\lambda_X}
	\left[
	\psi_{+1,\lambda_X}^{\lambda^\prime \ast}(x,\bm{k}_\perp)
	\psi_{+1,\lambda_X}^{\lambda}(x,\bm{k}_\perp)
	+
	\psi_{-1,\lambda_X}^{\lambda^\prime \ast}(x,\bm{k}_\perp)
	\psi_{-1,\lambda_X}^{\lambda}(x,\bm{k}_\perp)
	\right]\label{eq:T++},
\end{align}	
with $\bm{k}^\prime_\perp = \bm{k}_\perp + (1-x)\bm{q}_\perp$.
As shown in Eq.~(\ref{eq:M3}), the GFF $\bar{C}_g(q^2)$ is computed from the nonconservation of the partial EMT~\cite{Lorce:2018egm}, and this expression can be expanded as
\begin{align}
	-i q_\perp^{(1)} q_\perp^{(2)} M \bar{C}_g(q^2)
	= \frac{(\bm{q}_\perp)^2}{2p^+} \left(\mathcal{M}_{\uparrow\downarrow}^{+1} + \mathcal{M}_{\downarrow\uparrow}^{+1}\right)
	- q_\perp^{(1)} \left(\mathcal{M}_{\uparrow\downarrow}^{11} + \mathcal{M}_{\downarrow\uparrow}^{11}\right)
	- q_\perp^{(2)} \left(\mathcal{M}_{\uparrow\downarrow}^{21} + \mathcal{M}_{\downarrow\uparrow}^{21}\right).
\end{align}	
Due to the conservation of EMT in QCD, $\partial_\mu T^{\mu\nu}=0$, the total $\bar{C}(t)$ obtained by summing over all constituents should naturally vanish for all values of $t$, i.e., $\sum_{i=q,g}\bar{C}_i(t)=0$, and this is also independent of the renormalization scale.
In the overlap representation of LFWFs, the nucleon matrix elements of $T_g^{+1}$, $T_g^{11}$, and $T_g^{21}$ are given by~\cite{Sain:2025kup}
\begin{align}
	\langle p' &,\lambda' | T_g^{+1} | p,\lambda \rangle\nonumber\\
	=& p^+ \int \frac{\mathrm{d}^2 \bm{k}_\perp \mathrm{d}x}{16\pi^3}
	\sum_{\lambda_X} \bigg[
	\left(2k_\perp^{(1)} + q_\perp^{(1)} - i q_\perp^{(2)}\right)	\psi_{+1,\lambda_X}^{*\lambda'}(x,\bm{k}'_\perp)
\psi_{+1,\lambda_X}^{\lambda}(x,\bm{k}_\perp) \nonumber\\
	& + \left(2k_\perp^{(1)} + q_\perp^{(1)} + i q_\perp^{(2)}\right)	\psi_{-1,\lambda_X}^{*\lambda'}(x,\bm{k}'_\perp)
\psi_{-1,\lambda_X}^{\lambda}(x,\bm{k}_\perp)
	\bigg],\label{eq:T+1}\\
	\langle p' &,\lambda' | T_g^{11} | p,\lambda \rangle\nonumber\\
	=& \int \frac{\mathrm{d}^2 \bm{k}_\perp \mathrm{d}x}{16\pi^3 x}
	\sum_{\lambda_X} \bigg[
	\left(2(k_\perp^{(1)})^2 + 2k_\perp^{(1)}q_\perp^{(1)} + i k_\perp^{(2)}q_\perp^{(1)} - i k_\perp^{(1)}q_\perp^{(2)}\right)	\psi_{+1,\lambda_X}^{*\lambda'}(x,\bm{k}'_\perp)
\psi_{+1,\lambda_X}^{\lambda}(x,\bm{k}_\perp)\nonumber \\
	& + \left(2(k_\perp^{(1)})^2 + 2k_\perp^{(1)}q_\perp^{(1)} - i k_\perp^{(2)}q_\perp^{(1)} + i k_\perp^{(1)}q_\perp^{(2)}\right)	\psi_{-1,\lambda_X}^{*\lambda'}(x,\bm{k}'_\perp)
\psi_{-1,\lambda_X}^{\lambda}(x,\bm{k}_\perp)\nonumber \\
	& - \left(\frac{(\bm{q}_\perp)^2}{2} + i k_\perp^{(2)}q_\perp^{(1)} - i k_\perp^{(1)}q_\perp^{(2)}\right)	\left(\psi_{+1,\lambda_X}^{*\lambda'}(x,\bm{k}'_\perp)
\psi_{-1,\lambda_X}^{\lambda}(x,\bm{k}_\perp)
	+ \psi_{-1,\lambda_X}^{*\lambda'}(x,\bm{k}'_\perp)
\psi_{+1,\lambda_X}^{\lambda}(x,\bm{k}_\perp)\right)
	\bigg],\label{eq:T11}\\
	\langle p' &,\lambda' | T_g^{21} | p,\lambda \rangle\nonumber\\
	=& \int \frac{\mathrm{d}^2 \bm{k}_\perp \mathrm{d}x}{16\pi^3 x}
	\sum_{\lambda_X} \bigg[
	\left(2k_\perp^{(1)}k_\perp^{(2)} + k_\perp^{(2)}q_\perp^{(1)} + k_\perp^{(1)}q_\perp^{(2)}\right)	\left(\psi_{+1,\lambda_X}^{*\lambda'}(x,\bm{k}'_\perp)
\psi_{+1,\lambda_X}^{\lambda}(x,\bm{k}_\perp)
	+ \psi_{-1,\lambda_X}^{*\lambda'}(x,\bm{k}'_\perp)
\psi_{-1,\lambda_X}^{\lambda}(x,\bm{k}_\perp)\right)\nonumber \\
	& + i\left(\frac{(\bm{q}_\perp)^2}{2} + i k_\perp^{(2)}q_\perp^{(1)} - i k_\perp^{(1)}q_\perp^{(2)}\right)	\left(\psi_{+1,\lambda_X}^{*\lambda'}(x,\bm{k}'_\perp)
\psi_{-1,\lambda_X}^{\lambda}(x,\bm{k}_\perp)
	- \psi_{-1,\lambda_X}^{*\lambda'}(x,\bm{k}'_\perp)
\psi_{+1,\lambda_X}^{\lambda}(x,\bm{k}_\perp)\right)
	\bigg]\label{eq:T21},
\end{align}	
respectively.
Finally, the GFF $D_g(q^2)$ in Eq.~(\ref{eq:M4}) is derived from the nucleon matrix elements of $T_g^{11}$ and $T_g^{22}$, and their sum can be written in the overlap representation of LFWFs as~\cite{Sain:2025kup}
\begin{align}
	\langle p' &,\lambda' | (T_g^{11}+T_g^{22}) | p,\lambda \rangle\nonumber\\
	=& 2\int \frac{\mathrm{d}^2 \bm{k}_\perp \mathrm{d}x}{16\pi^3 x}
	\sum_{\lambda_X} \bigg[
	\left((k_\perp^{(1)})^2+(k_\perp^{(2)})^2 + k_\perp^{(1)}q_\perp^{(1)}+k_\perp^{(2)}q_\perp^{(2)}\right)	\Big(\psi_{+1,\lambda_X}^{*\lambda'}(x,\bm{k}_\perp')
\psi_{+1,\lambda_X}^{\lambda}(x,\bm{k}_\perp) \nonumber\\
	& + \psi_{-1,\lambda_X}^{*\lambda'}(x,\bm{k}_\perp')
\psi_{-1,\lambda_X}^{\lambda}(x,\bm{k}_\perp)\Big)
	+ i\left(k_\perp^{(2)}q_\perp^{(1)}-k_\perp^{(1)}q_\perp^{(2)}\right)	\Big(\psi_{+1,\lambda_X}^{*\lambda'}(x,\bm{k}_\perp')
\psi_{+1,\lambda_X}^{\lambda}(x,\bm{k}_\perp)\nonumber \\
	& - \psi_{-1,\lambda_X}^{*\lambda'}(x,\bm{k}_\perp')
\psi_{-1,\lambda_X}^{\lambda}(x,\bm{k}_\perp)\Big)
	\bigg]\label{eq:T11T22}.
\end{align}	

Using the LFWFs in Eqs.~(\ref{eq:wavefunction+}-\ref{eq:wavefunction-}), we evaluate the matrix elements in Eqs.~(\ref{eq:T++}) and (\ref{eq:T+1}-\ref{eq:T11T22}), and extract the gluon GFFs $A_g(q^2)$, $B_g(q^2)$, $D_g(q^2)$, and $\bar{C}_g(q^2)$ from Eqs.~(\ref{eq:M1}-\ref{eq:M4}). 
The corresponding analytical expressions are as follows:
\begin{align}
	A_g(q_\perp^2)=&\int_0^1 dx\int_0^\infty dk_\perp \frac{ N_g^2 k_\perp}{8\pi^3 \{k_\perp^2 + [M_X^2 + M^2 (-1 + x)] x\}} \exp\left( \frac{2 k_\perp^2 + q_\perp^2 (-1 + x)^2 + 2 M_X^2 x}{2 (-1 + x) x \beta_1^2} \right) \nonumber\\
	&\times \bigg\{ \pi [2 + (-2 + x) x] I_0\left( \frac{k_\perp q_\perp}{x \beta_1^2} \right) + \frac{1}{\sqrt{ -4 k_\perp^2 q_\perp^2 (1 - x)^2 + [ k_\perp^2 + q_\perp^2 (1 - x)^2 + (M_X^2 + M^2 (-1 + x)) x ]^2 }}\nonumber\\
	&\times\bigg[ \pi ( k_\perp^2 (2 + (-2 + x) x) - q_\perp^2 (-1 + x)^2 (2 + (-2 + x) x) + x ( 4 M M_X (-1 + x) x - M_X^2 (2 + (-4 + x) x) \nonumber\\
	&+ M^2 (1 - x) (2 - x^2) ) ) \bigg( I_0\left( \frac{k_\perp q_\perp}{x \beta_1^2} \right) + 2 \sum_{n=1}^\infty 2^{-n} \bigg( \frac{1}{k_\perp q_\perp (1 - x)} \Big( k_\perp^2 + q_\perp^2 (1-x)^2 + (M_X^2 - M^2)x  + M^2 x^2 \nonumber\\
	&- \sqrt{( k_\perp^2 + q_\perp^2 - M^2 x + M_X^2 x - 2 q_\perp^2 x + M^2 x^2 + q_\perp^2 x^2 )^2 -4 k_\perp^2 q_\perp^2 (1 - x)^2  } \Big) \bigg)^n I_n\left( \frac{k_\perp q_\perp}{x \beta_1^2} \right) \bigg) \bigg] \bigg\},\\
	B_g(q_\perp^2)=&\int_0^1 dx\int_0^\infty dk_\perp \frac{ N_g^2 k_\perp M [M_X + M (-1 + x)] (-1 + x)^2 x}{2 \pi^2 \{ k_\perp^2 + [ M_X^2 + M^2 (-1 + x) ] x \} } \exp\left( \frac{2 k_\perp^2 + q_\perp^2 (-1 + x)^2 + 2 M_X^2 x}{2 (-1 + x) x \beta^2} \right)\nonumber\\
	&\times \frac{1}{\sqrt{ -4 k_\perp^2 q_\perp^2 (-1 + x)^2 + \{ k_\perp^2 + q_\perp^2 (-1 + x)^2 + [ M_X^2 + M^2 (-1 + x) ] x \}^2 }} \nonumber\\
	& \times \bigg\{ I_0\left( \frac{k_\perp q_\perp}{x \beta^2} \right)+ 2 \sum_{n=1}^\infty \left(-\frac{1}{2}\right)^n \bigg[ \frac{1}{ k_\perp q_\perp (1 - x) }\Big(k_\perp^2 + q_\perp^2(1-x)^2 - M^2 x (1-x) + M_X^2 x\nonumber
\end{align}
\begin{align}	
	& - \sqrt{( k_\perp^2 + q_\perp^2(1-x)^2 - M^2 x(1-x) + M_X^2 x )^2 -4 k_\perp^2 q_\perp^2 (1 - x)^2  } \Big) \bigg]^n I_n\left( \frac{k_\perp q_\perp}{x \beta^2} \right) \bigg\},\\
	D_g(q_\perp^2)=&\int_0^1 dx\int_0^\infty dk_\perp \frac{ N_g^2 k_\perp M [M_X - M (1 - x)] (-1 + x)}{ 6 \pi^2 q_\perp^2 x \{k_\perp^2 + [M_X^2 - M^2 (1 - x)] x\} } \exp\left(\frac{2 k_\perp^2 + q_\perp^2 (-1 + x)^2 + 2 M_X^2 x}{2 (-1 + x) x \beta^2}\right) \nonumber\\
	&\times \bigg\{ \bigg[ \frac{2 (k_\perp^2 + (M_X^2 - q_\perp^2 - M^2 (1 - x)) x)}{q_\perp^2 (1 - x)} + \frac{1}{ \sqrt{ (k_\perp^2 + q_\perp^2 (1 - x)^2 + (M_X^2 - M^2 (1 - x)) x)^2 -4 k_\perp^2 q_\perp^2 (1 - x)^2 } }\nonumber\\	
	&\times \left( \frac{2 (k_\perp^2 + q_\perp^2 (1 - x)^2 + (M_X^2 - M^2 (1 - x)) x) (k_\perp^2 + (M_X^2 - q_\perp^2 - M^2 (1 - x)) x)}{q_\perp^2 (-1 + x)} + (1 - x) (4 k_\perp^2 - q_\perp^2 x^2) \right) \bigg]\nonumber\\
	&\times I_0\left( \frac{k_\perp q_\perp}{x \beta^2} \right) + \frac{4 k_\perp I_1\left( \frac{k_\perp q_\perp}{x \beta^2} \right)}{q_\perp} + \frac{1}{ \sqrt{ [k_\perp^2 + q_\perp^2 (1 - x)^2 + (M_X^2 + M^2 (-1 + x)) x]^2 -4 k_\perp^2 q_\perp^2 (1 - x)^2 } } \nonumber\\
	&\times\bigg[ 2 \bigg( \frac{2 (k_\perp^2 + q_\perp^2 (1 - x)^2 + (M_X^2 - M^2 (1 - x)) x) (k_\perp^2 + (M_X^2 - q_\perp^2 - M^2 (1 - x)) x)}{q_\perp^2 (-1 + x)} \nonumber\\
	& + (1 - x) (4 k_\perp^2 - q_\perp^2 x^2) \bigg) \sum_{n=1}^\infty 2^{-n} \bigg( \frac{1}{ k_\perp q_\perp (1 - x) } \Big(k_\perp^2 + q_\perp^2 (1 - x)^2 - (M^2 - M_X^2) x + M^2 x^2 \nonumber\\
	&- \sqrt{ (k_\perp^2 + q_\perp^2 (1 - x)^2 - (M^2 - M_X^2) x + M^2 x^2)^2 -4 k_\perp^2 q_\perp^2 (1 - x)^2 } \Big) \bigg)^n I_n\left( \frac{k_\perp q_\perp}{x \beta^2} \right) \bigg] \bigg\},\\
	\bar{C}_g(q_\perp^2)=&\int_0^1 dx\int_0^\infty dk_\perp \frac{  N_g^2k_\perp^2 [M_X + M (-1 + x)] (-1 + x)}{ 4 M \pi^2 x \{k_\perp^2 + [M_X^2 + M^2 (-1 + x)] x\} } \exp\left(\frac{2 k_\perp^2 + q_\perp^2 (-1 + x)^2 + 2 M_X^2 x}{2 (-1 + x) x \beta^2}\right)\nonumber\\
	&\times \Bigg\{ \Bigg[ -\frac{(-2 + x) (k_\perp^2 + q_\perp^2 (-1 + x)^2 + (M_X^2 + M^2 (-1 + x)) x)}{2 k_\perp q_\perp^2 (-1 + x)^2} + \frac{2 + (-2 + x) x}{2 k_\perp (-1 + x)} \nonumber\\
	&+ \frac{ k_\perp x - \frac{(k_\perp^2 + q_\perp^2 (1 - x)^2 + (M_X^2 - M^2 (1 - x)) x) (2 - (2 - x) x)}{2 k_\perp (-1 + x)} + k_\perp (2 - x) \Big( 1 - \frac{(k_\perp^2 + q_\perp^2 (1 - x)^2 + (M_X^2 + M^2 (-1 + x)) x)^2}{2 k_\perp^2 q_\perp^2 (-1 + x)^2} \Big) }{ \sqrt{ -4 k_\perp^2 q_\perp^2 (-1 + x)^2 + (k_\perp^2 + q_\perp^2 (-1 + x)^2 + (M_X^2 + M^2 (-1 + x)) x)^2 } } \Bigg]\nonumber\\
	&\times I_0\left( \frac{k_\perp q_\perp}{x \beta^2} \right) + \frac{(2 - x) I_1\left( \frac{k_\perp q_\perp}{x \beta^2} \right)}{q_\perp (1 - x)} + \frac{1}{ \sqrt{ [k_\perp^2 + q_\perp^2 (1 - x)^2 + (M_X^2 + M^2 (-1 + x)) x]^2 -4 k_\perp^2 q_\perp^2 (1 - x)^2 } }\nonumber\\
	&\times \bigg[2 \bigg( k_\perp x - \frac{(k_\perp^2 + q_\perp^2 (-1 + x)^2 + (M_X^2 + M^2 (-1 + x)) x) (2 + (-2 + x) x)}{2 k_\perp (-1 + x)} + k_\perp (-2 + x)\nonumber\\
	&\times \left( \frac{(k_\perp^2 + q_\perp^2 (1 - x)^2 + (M_X^2 - M^2 (1 - x)) x)^2}{2 k_\perp^2 q_\perp^2 (1 - x)^2} -1 \right) \bigg) \sum_{n=1}^\infty 2^{-n} \bigg( \frac{1}{ k_\perp q_\perp (1 - x) } \Big( k_\perp^2 + q_\perp^2 (1 - x)^2 + M_X^2 x\nonumber\\
	& - M^2 (1 - x) x - \sqrt{ (k_\perp^2 + q_\perp^2 (1 - x)^2 + M_X^2 x - M^2 (1 - x) x)^2 -4 k_\perp^2 q_\perp^2 (1 - x)^2 } \Big) \bigg)^n I_n\left( \frac{k_\perp q_\perp}{x \beta^2} \right) \bigg] \Bigg\},
\end{align}	
where the modified Bessel function of the first kind $I_n(x)$ for integer order $n$ is defined by the series
\begin{align}
	I_n(x)=\sum_{k=0}^{\infty}\frac{1}{k!(k+n)!}\left(\frac{x}{2}\right)^{2k+n},
\end{align}	
and admits the integral representation 
\begin{align}
	I_n(x)=\frac{1}{\pi}\int_{0}^{\pi}d\theta e^{x\cos\theta}\cos(n\theta).
\end{align}

\section{NUMERICAL RESULTS AND DISCUSSION}\label{Sec:4}

\subsection{Parameter determination}

\begin{figure}
	\centering
	\includegraphics[width=0.3\columnwidth]{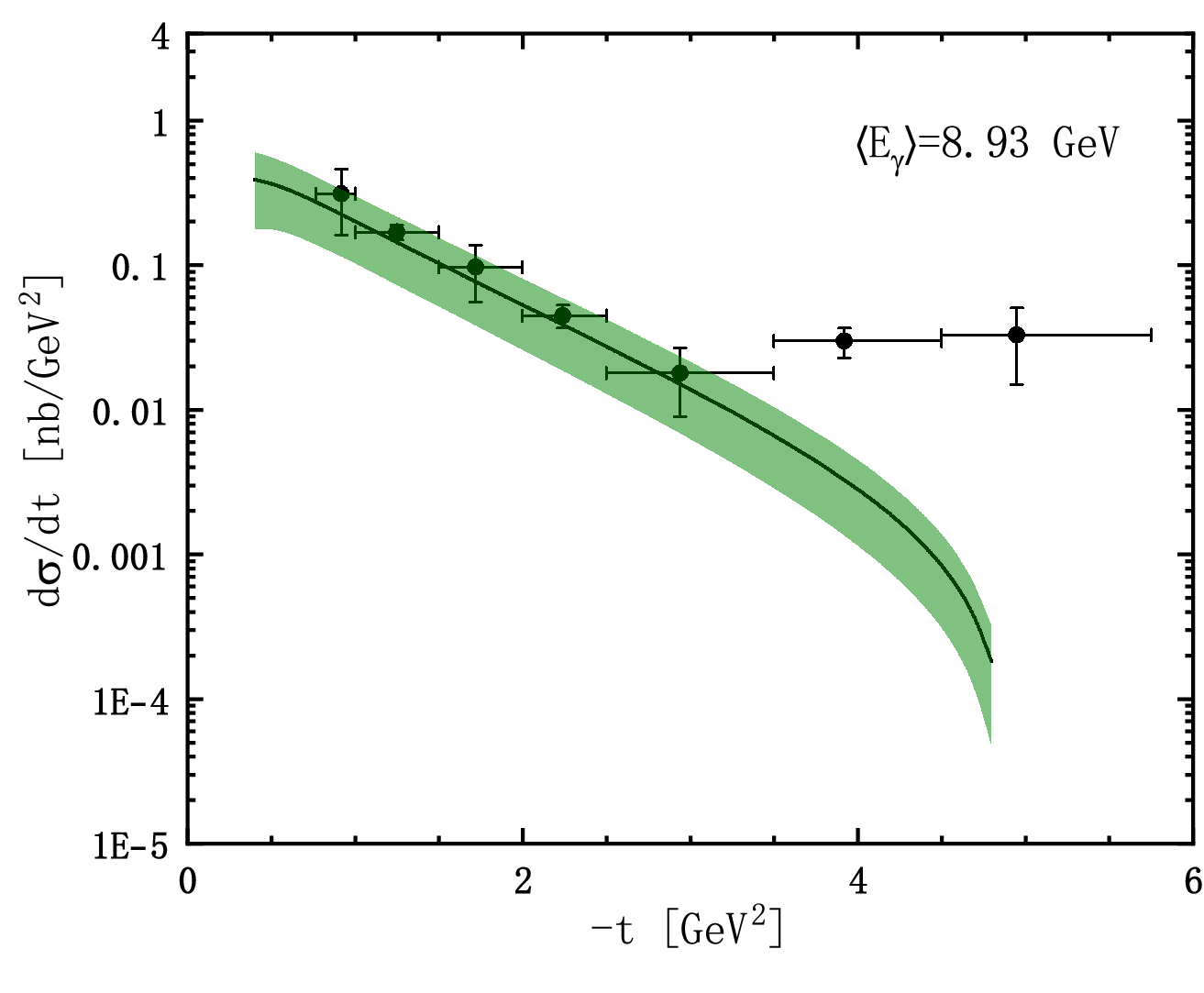}
	\includegraphics[width=0.3\columnwidth]{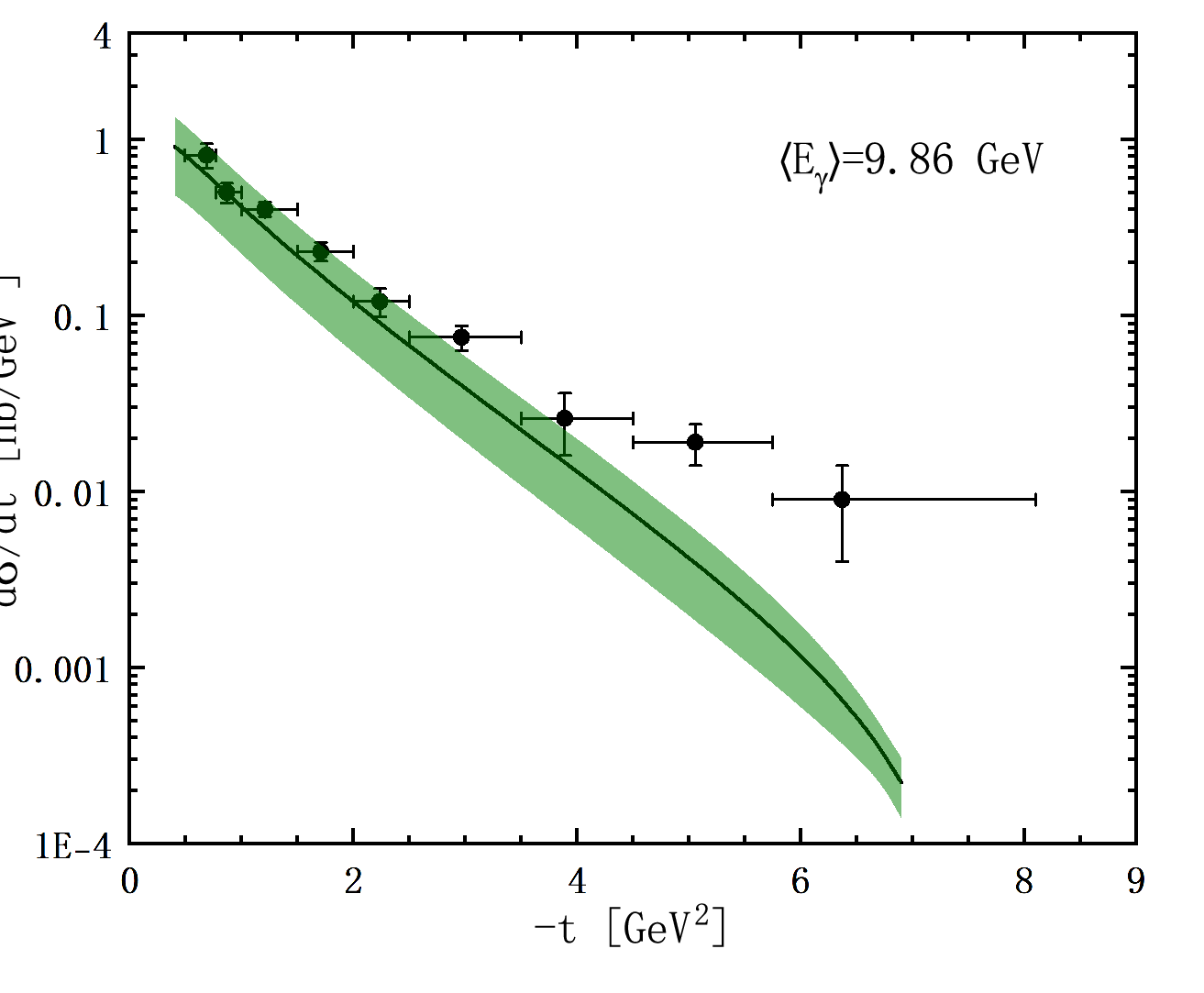}
	\includegraphics[width=0.3\columnwidth]{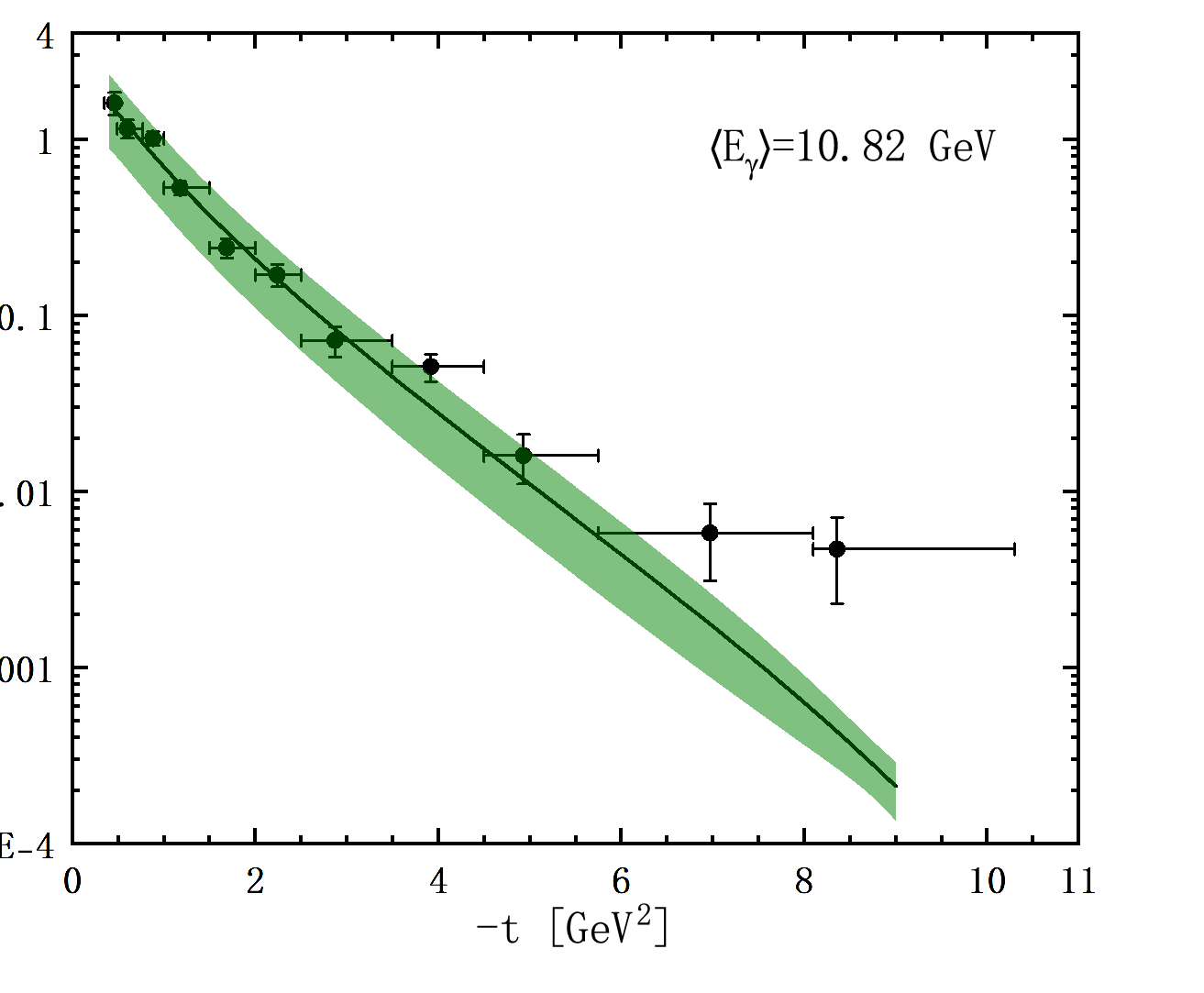}\\
	\includegraphics[width=0.3\columnwidth]{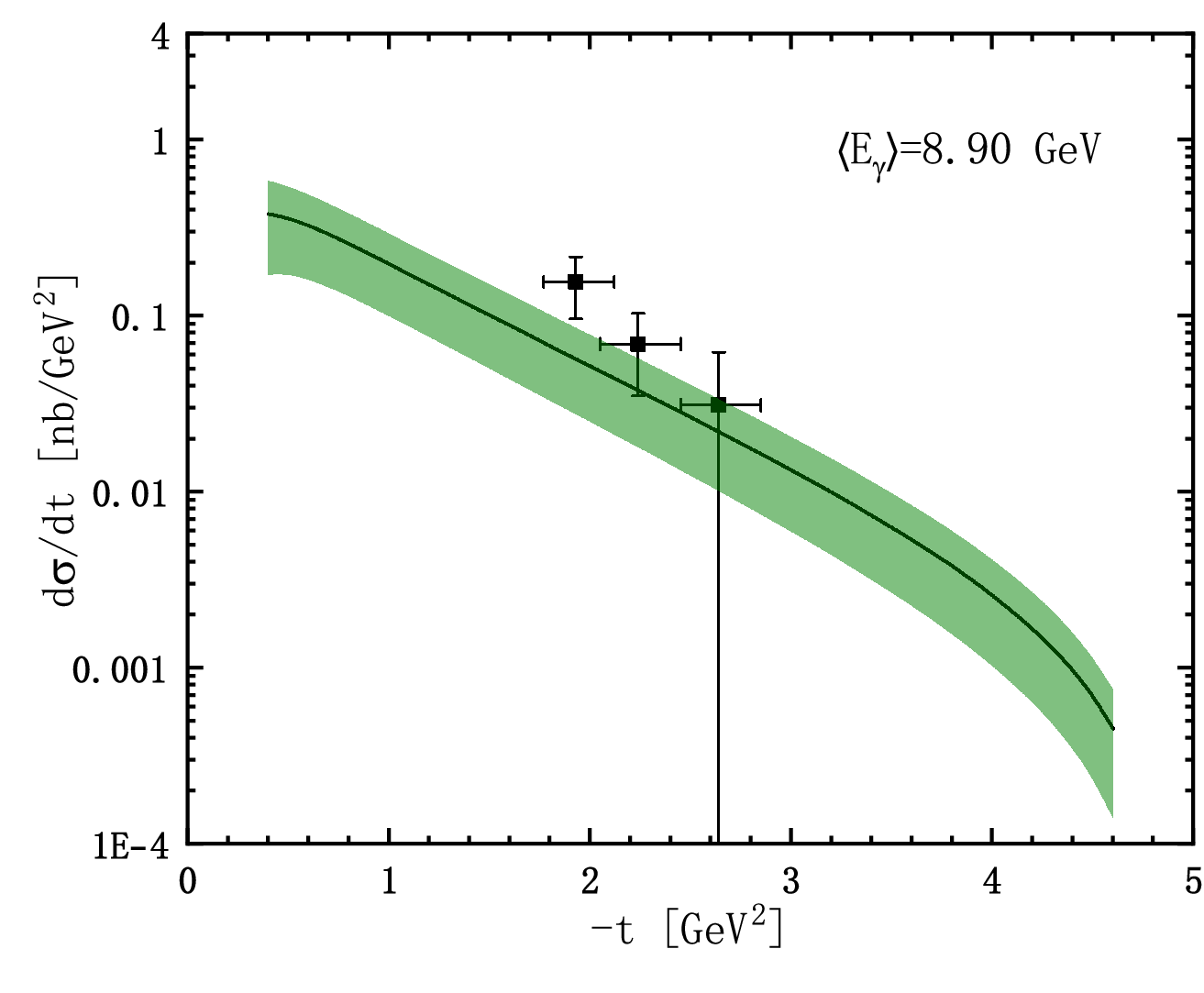} 
	\includegraphics[width=0.3\columnwidth]{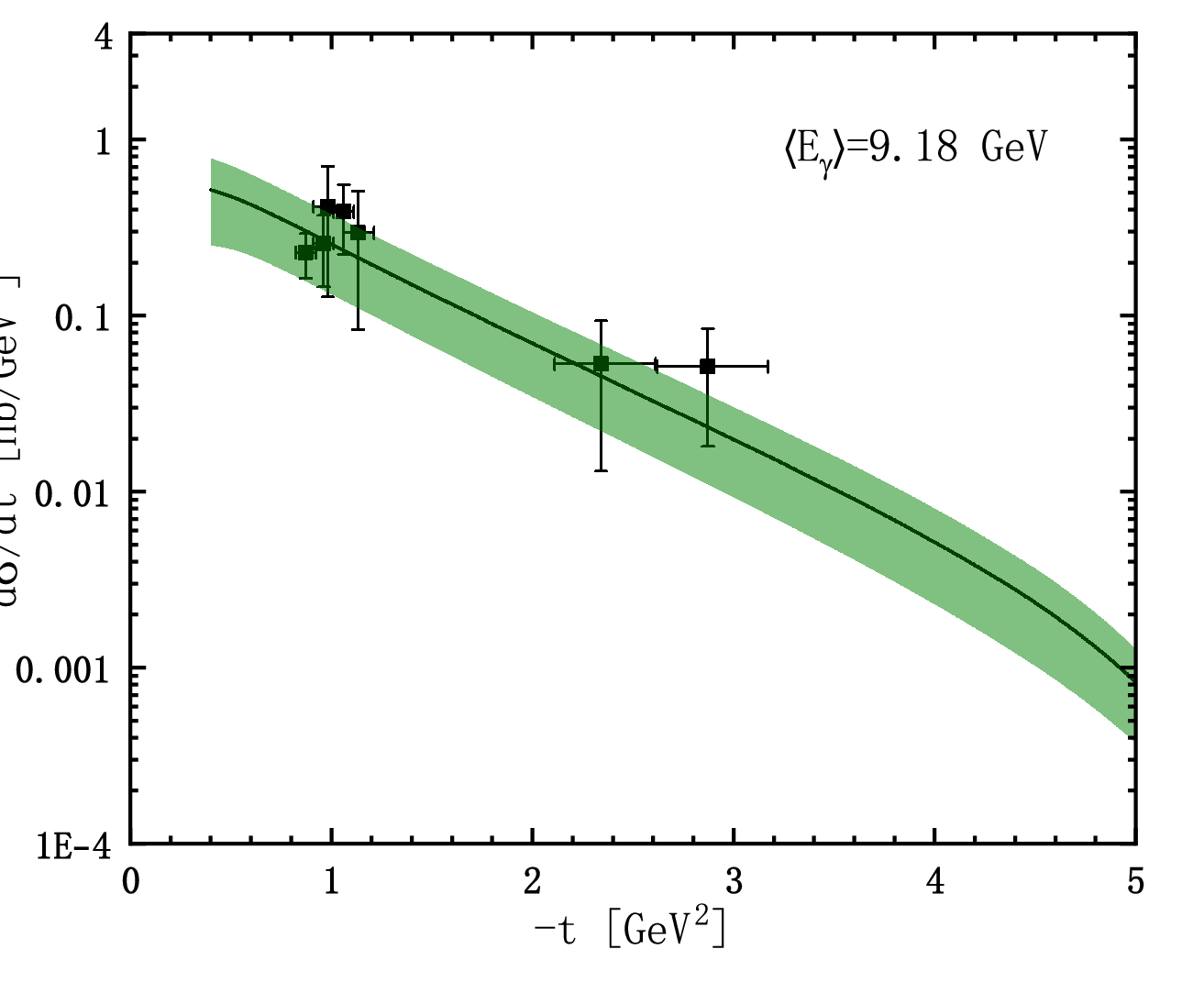}
	\includegraphics[width=0.3\columnwidth]{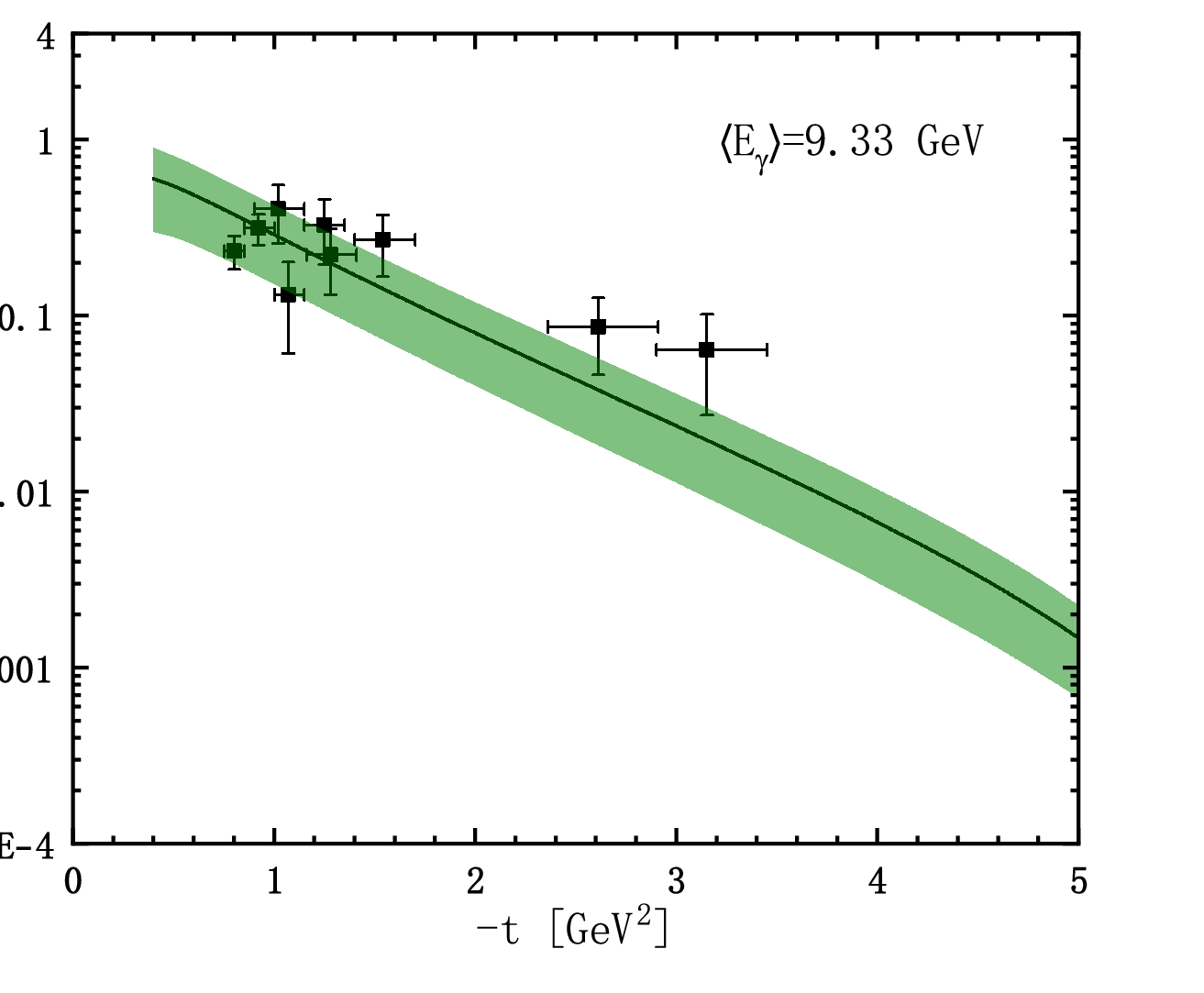}\\
	\includegraphics[width=0.3\columnwidth]{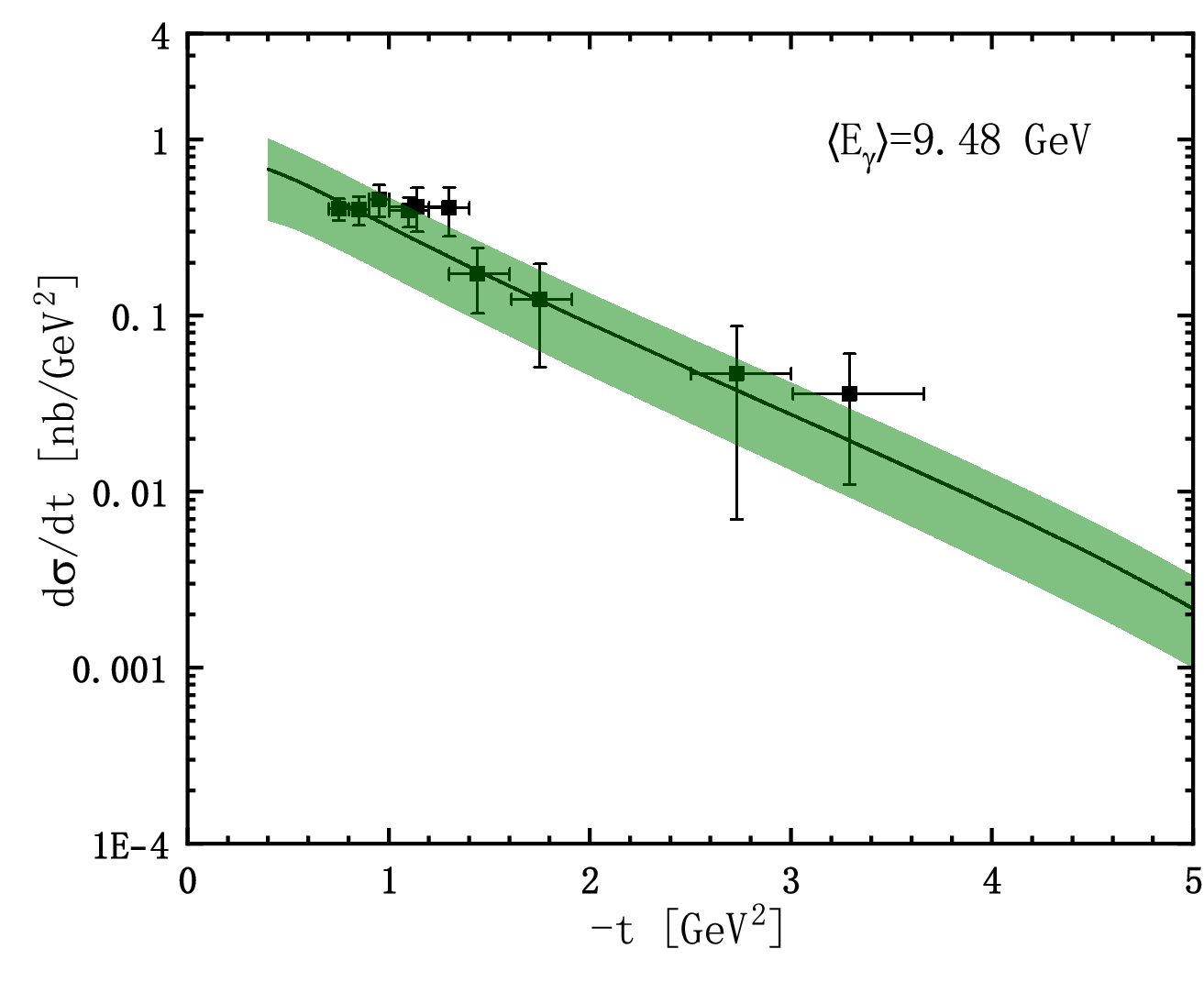}
	\includegraphics[width=0.3\columnwidth]{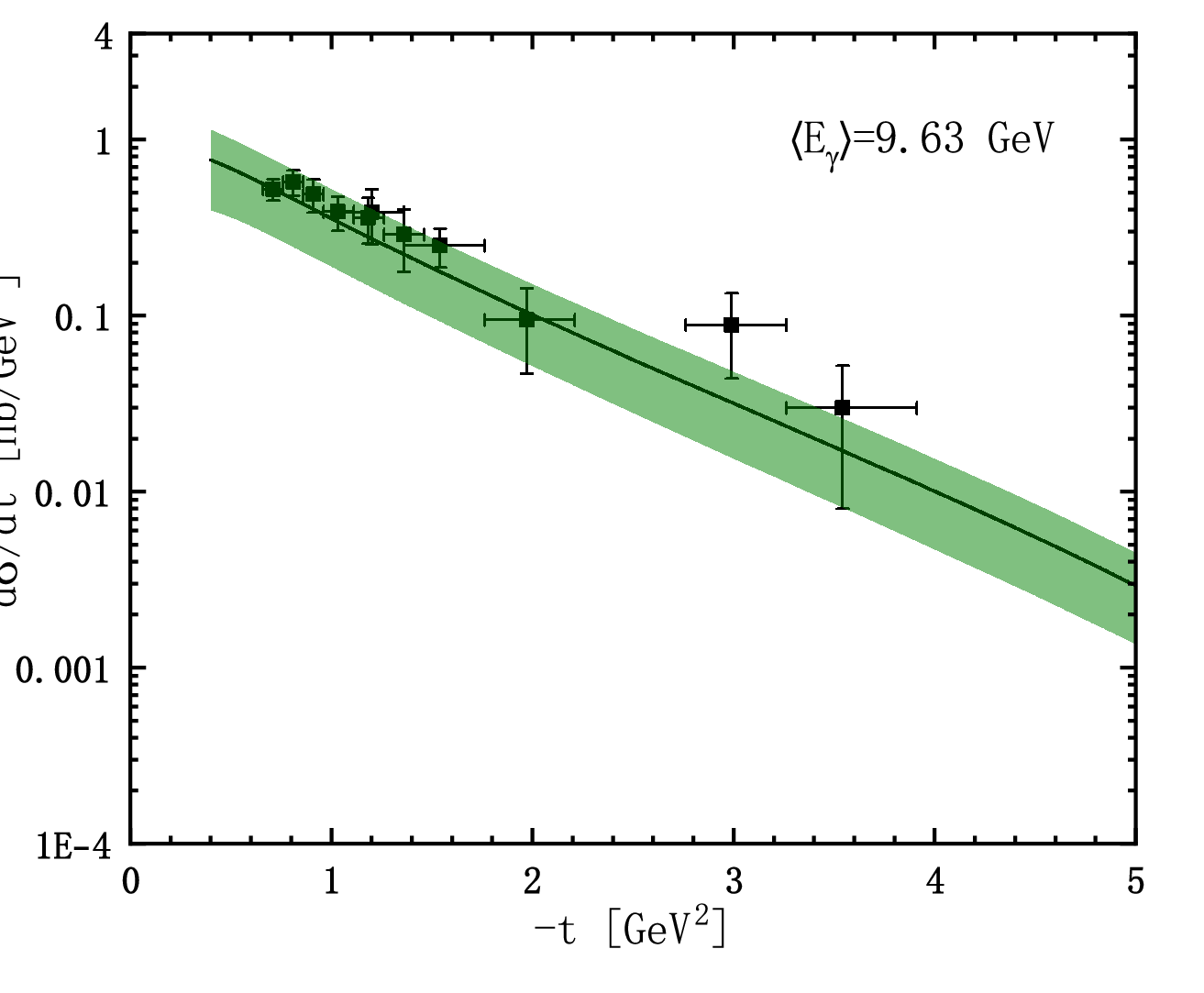}
	\includegraphics[width=0.3\columnwidth]{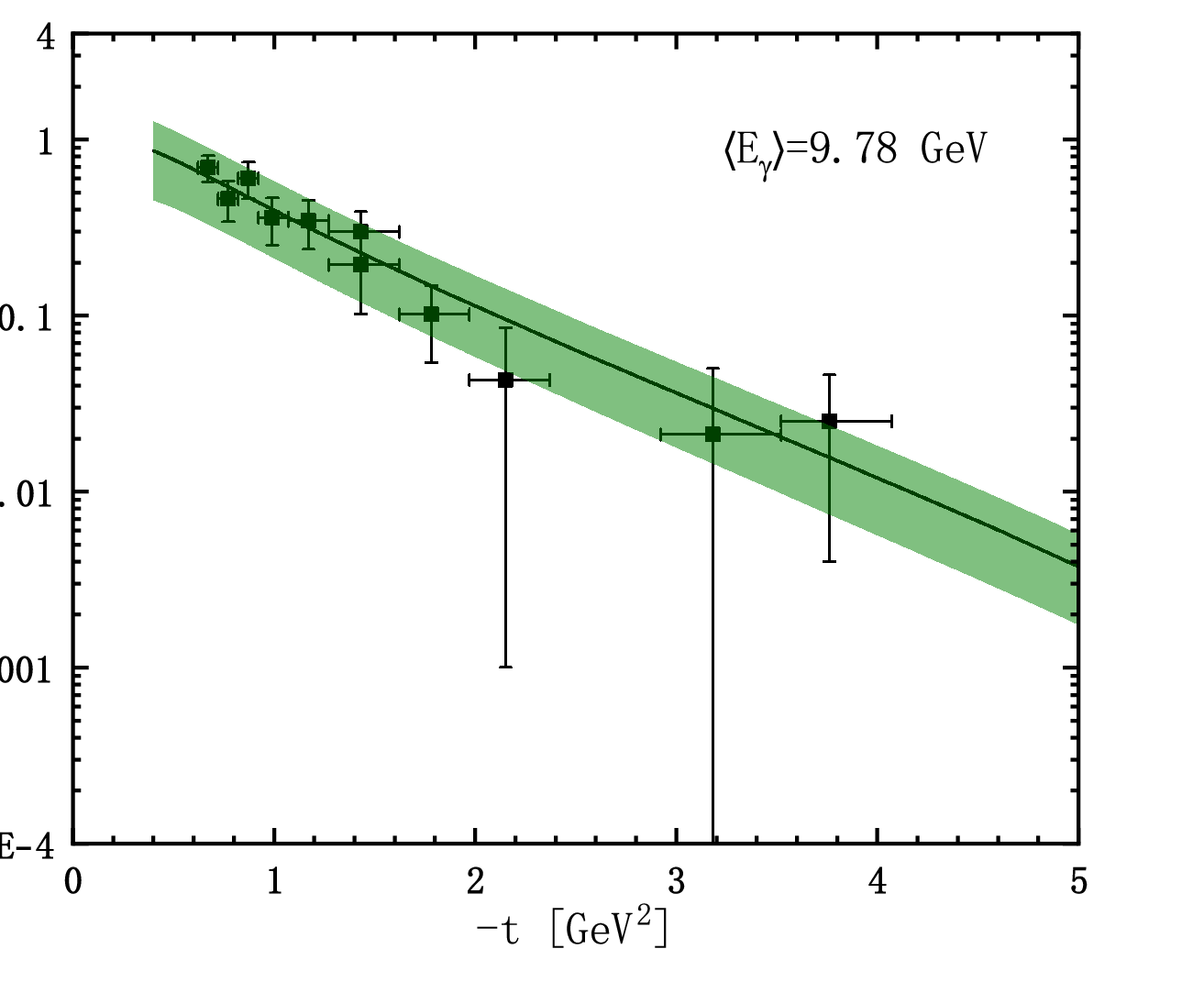}\\
	\includegraphics[width=0.3\columnwidth]{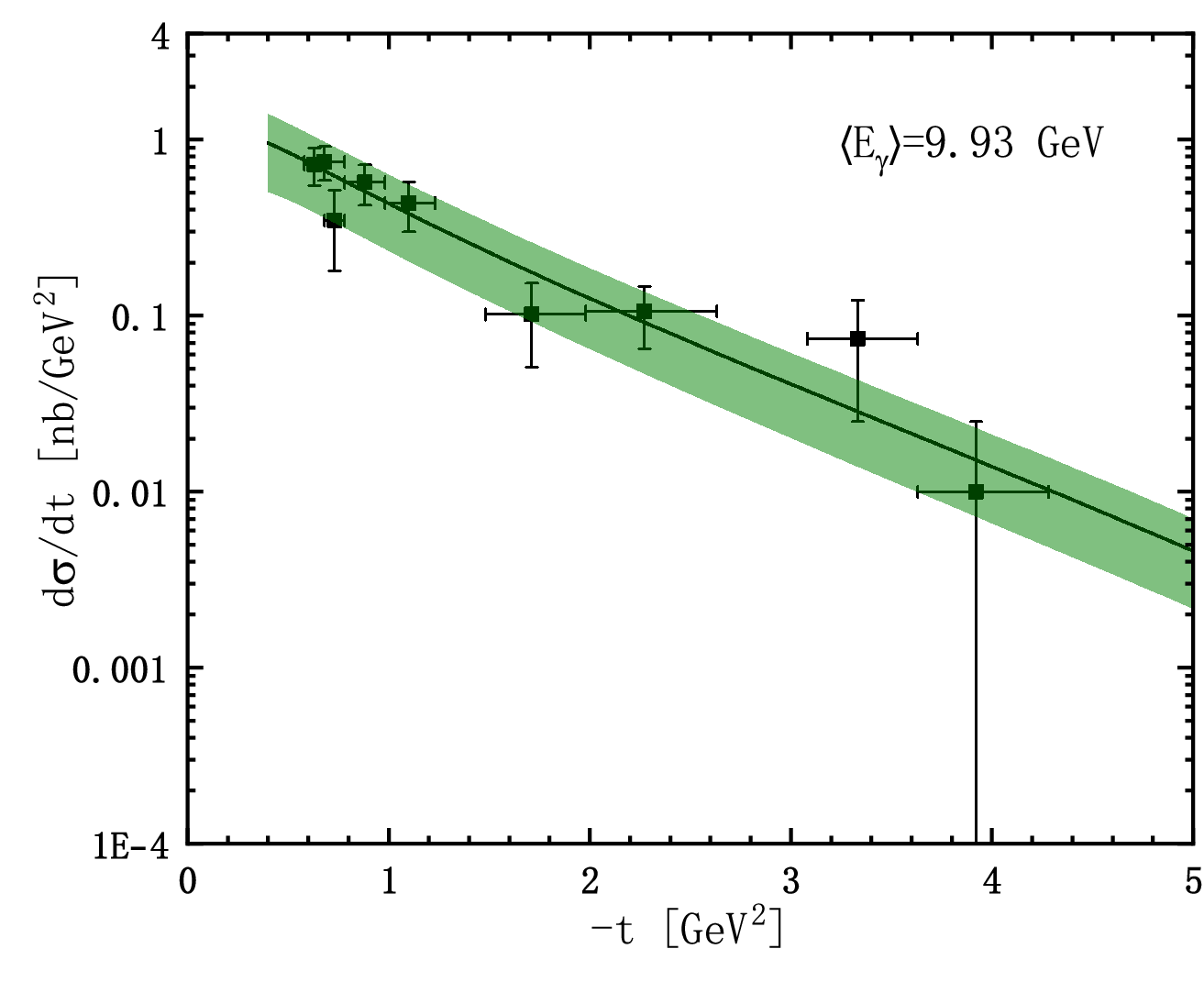}
	\includegraphics[width=0.3\columnwidth]{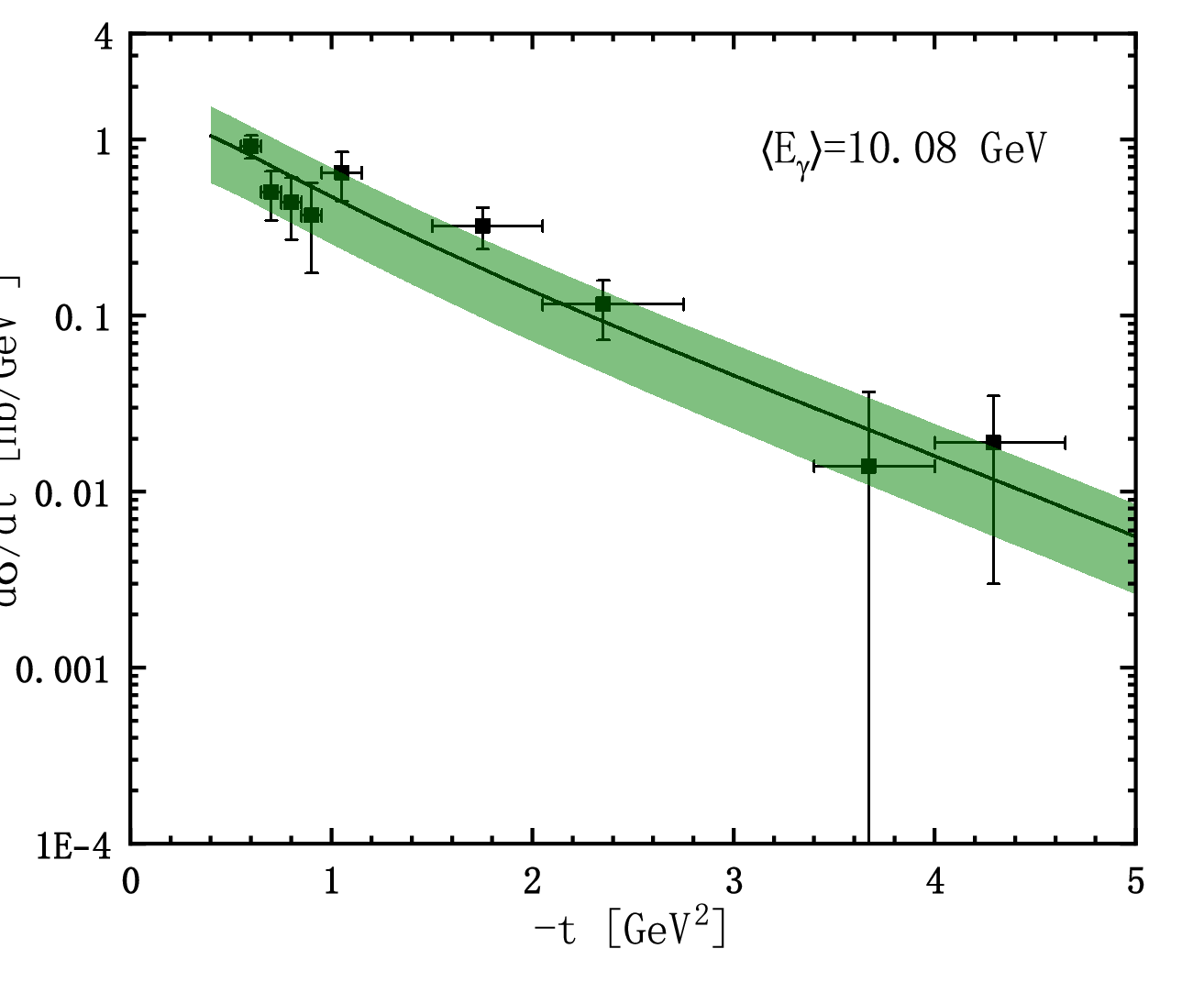}
	\includegraphics[width=0.3\columnwidth]{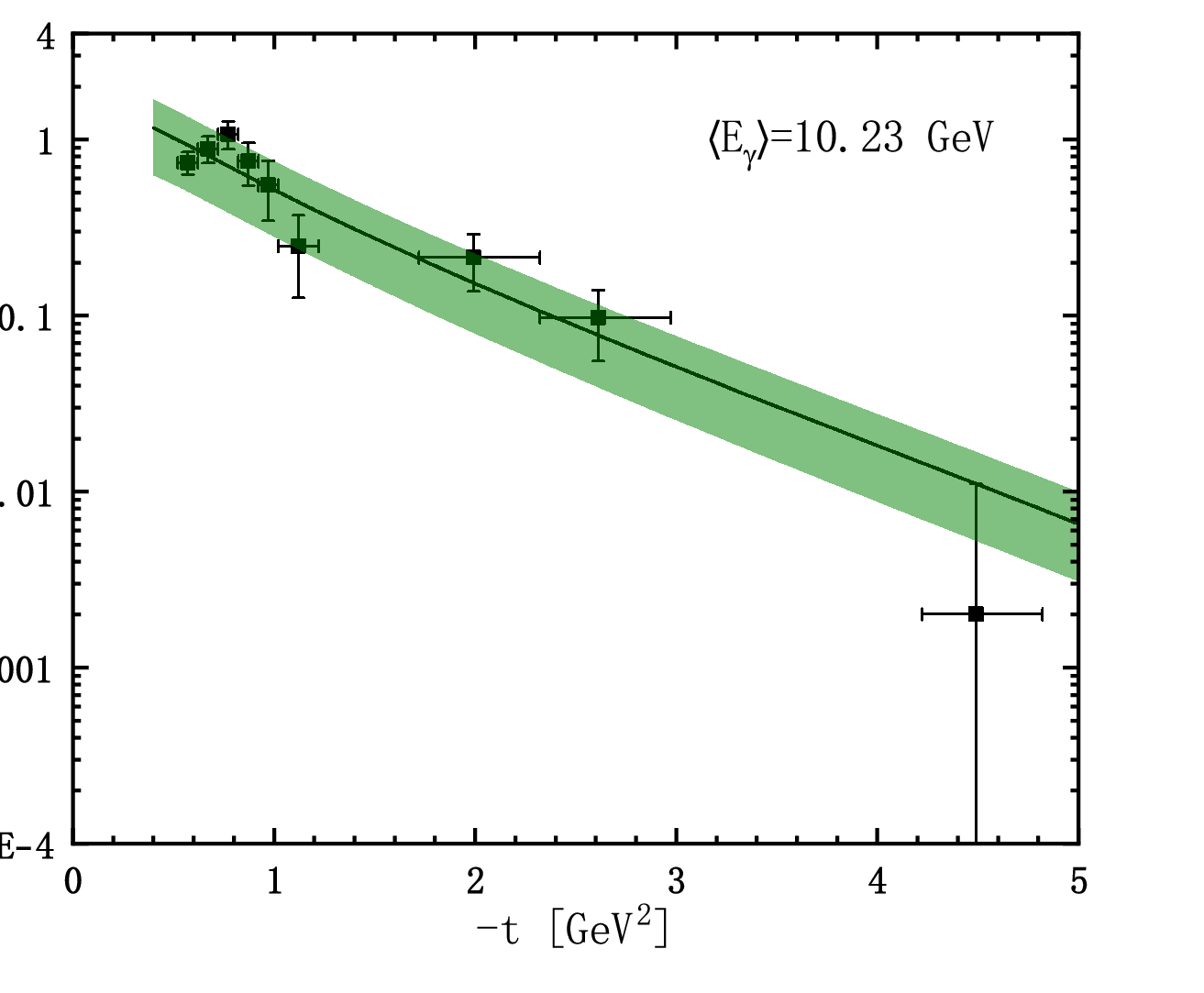}\\
	\includegraphics[width=0.3\columnwidth]{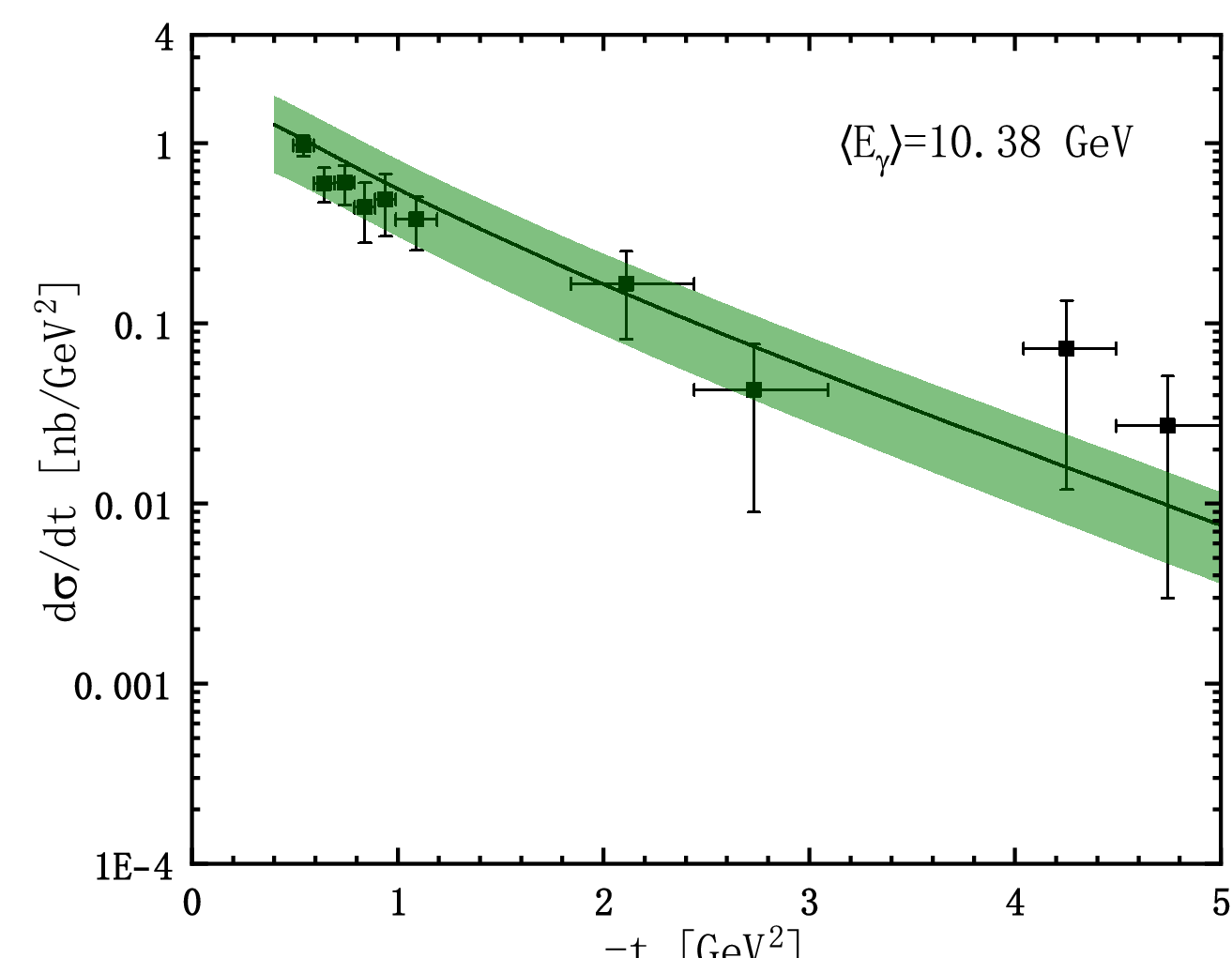}
	\includegraphics[width=0.3\columnwidth]{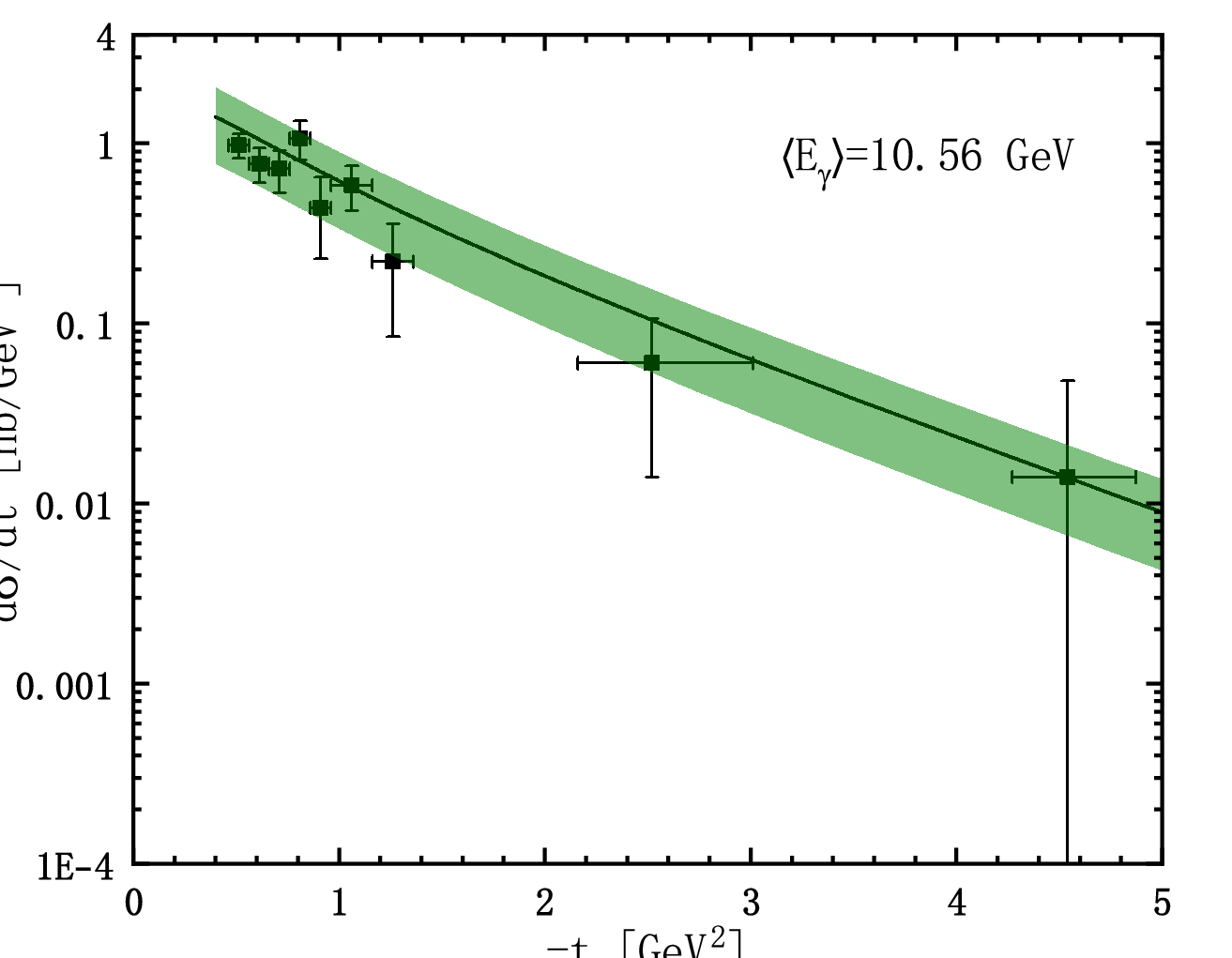}
	\includegraphics[width=0.3\columnwidth]{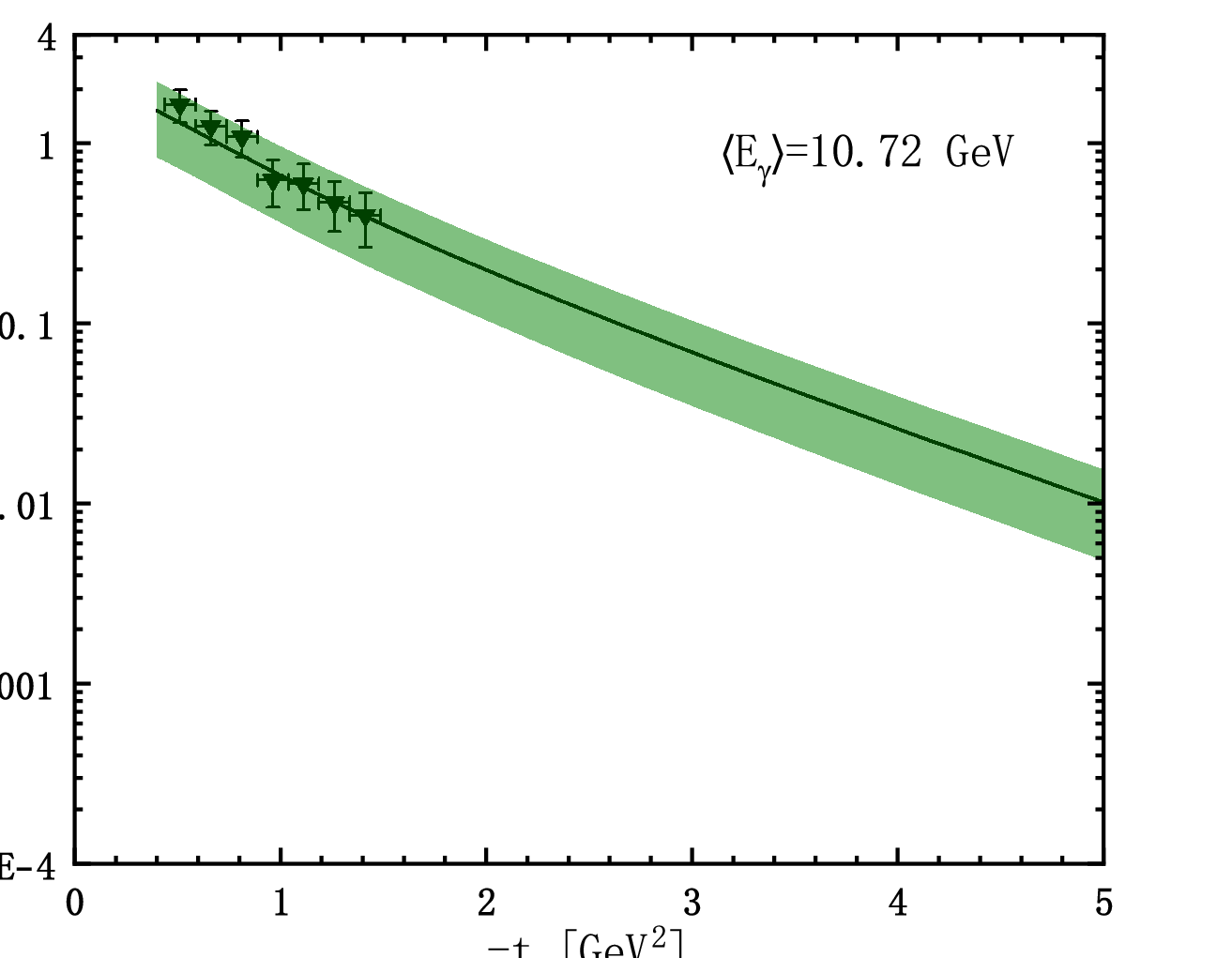}\\
	\caption{Global fit results for the $\gamma p \rightarrow J/\psi\, p$ differential cross section as a function of $-t$ at different $\langle E_\gamma\rangle$ values. The triangle, circle, and square data points correspond to the data from GlueX19~\cite{GlueX:2019mkq}, GlueX23~\cite{GlueX:2023pev}, and $J/\psi$-007~\cite{Duran:2022xag}, respectively, where the GlueX23 data span the entire physical $t$ range. The bands represent the parameter uncertainties.}
	\label{fig:dif}
\end{figure}

We first analyze all available data for the near-threshold $J/\psi$ photoproduction from the GlueX~\cite{GlueX:2019mkq,GlueX:2023pev} and $J/\psi$-007~\cite{Duran:2022xag} experiments. 
These data include differential cross sections in 15 photon-energy bins and total cross sections, with the GlueX23 differential cross section data covering the full near-threshold kinematic region. 
They have been well described in various theoretical studies on gluonic properties of the proton~\cite{Hatta:2019lxo,Mamo:2019mka,Mamo:2022eui,Sun:2021pyw,Sun:2021gmi,
Guo:2021ibg,Guo:2023pqw,Wang:2022ndz,Guo:2023qgu,Pentchev:2025qyn,Du:2020bqj,
JointPhysicsAnalysisCenter:2023qgg,Sakinah:2024cza,Tang:2024pky,Kim:2025oyo,
Lee:2022ymp,Kim:2024lxc,Mamo:2021krl,Kharzeev:2021qkd,Sain:2026ran,Tan:2026qba}. 
In this work, the differential cross section given by Eq.~(\ref{eq:dsigma}) is computed at different average photon energies $\langle E_\gamma\rangle$ using the light-front spectator model.
For the parameters related to the $J/\psi$ production~\cite{Guo:2021ibg}, we adopt $\alpha_\text{EM}=1/137$, $\alpha_S(\mu_0=2\,\text{GeV})=0.3$, $e_c=2/3$, $M=0.938\;\text{GeV}$, $M_{J/\psi}=3.097\;\text{GeV}$, and the $J/\psi$ wave function at the origin~\cite{Eichten:1995ch,Eichten:2019hbb}
\begin{align}
	|\psi_{\text{NR}}(0)|^2={1.0952\over 4\pi}(\text{GeV})^3.
\end{align}	

A global fit to the $t$-dependent differential cross section data for $J/\psi$ production at different $\langle E_\gamma\rangle$~\cite{GlueX:2019mkq,GlueX:2023pev,Duran:2022xag} determines the free parameters $N_\lambda$, $\beta_1$, and $M_X$, which are listed in Table~\ref{tab1}.
Fig.~\ref{fig:dif} compares the model results (solid curves) with the experimental data (solid circle points) in different $-t$ bins; the bands indicate the parameter uncertainties. 
Overall, the model describes the data well, with the main deviation from the GlueX23 results occurring at large $-t$.

\begin{table}[htbp]
	\centering
	\caption{Model parameters with uncertainties obtained by a global fit to the $\gamma p \rightarrow J/\psi\, p$ differential cross section experimental data~\cite{GlueX:2019mkq,GlueX:2023pev,Duran:2022xag}.}
	\label{tab1}
	\begin{tabular}{lcc} 
		\hline   
		Parameter & Central value & Uncertainty \\ 
		\hline 
		~~~~~$N_\lambda$ & 4.226  & $\pm0.326$    \\
		~~~~~$\beta_1$ & 3.253  & $\pm0.186$    \\
		~~~~~$M_X$ & 1.004  & $\pm0.029$   \\
		\hline 
	\end{tabular}
\end{table}

\subsection{Gluon GFFs and the total cross section}

\begin{figure}
	\centering
	\includegraphics[width=0.43\columnwidth]{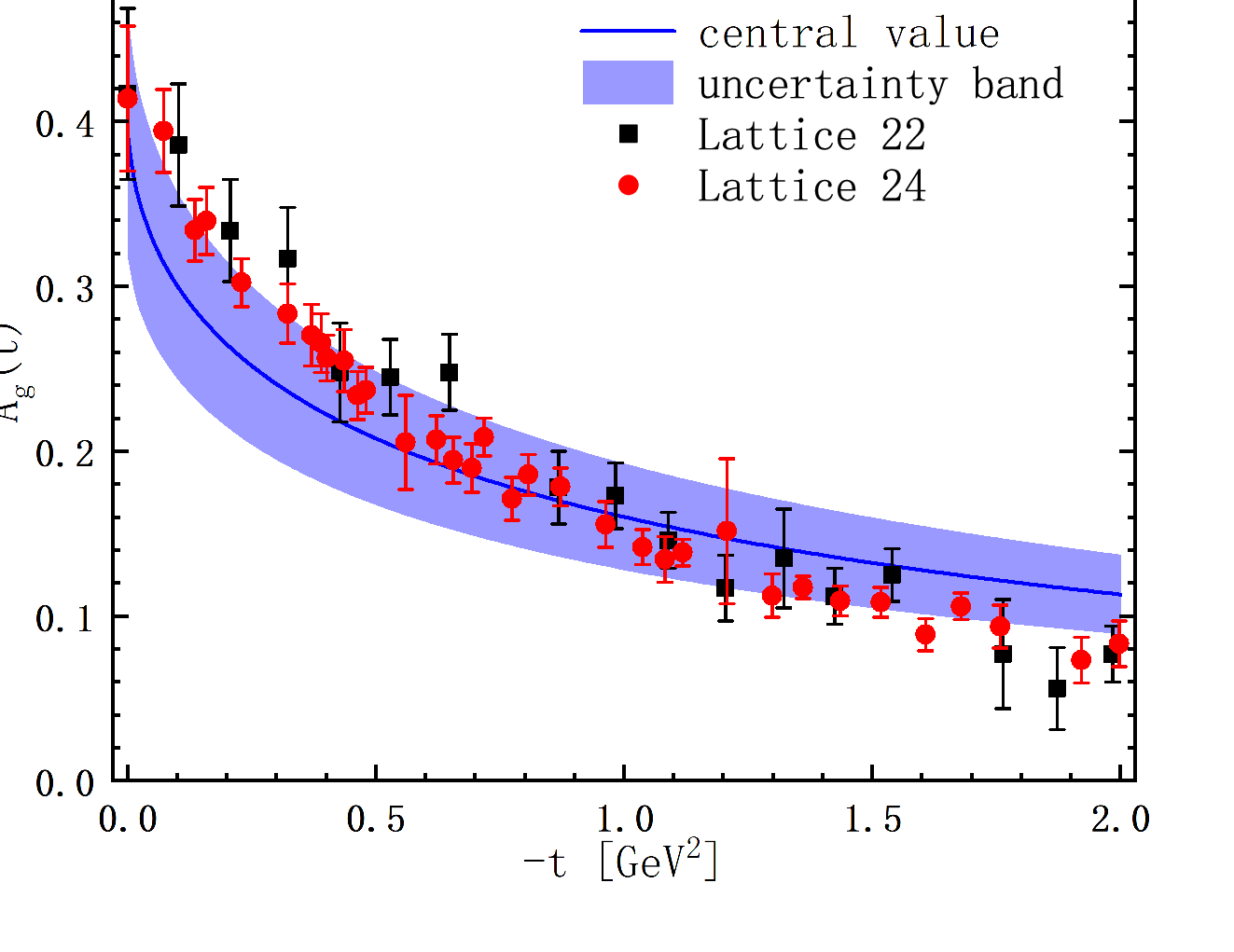}
	\includegraphics[width=0.43\columnwidth]{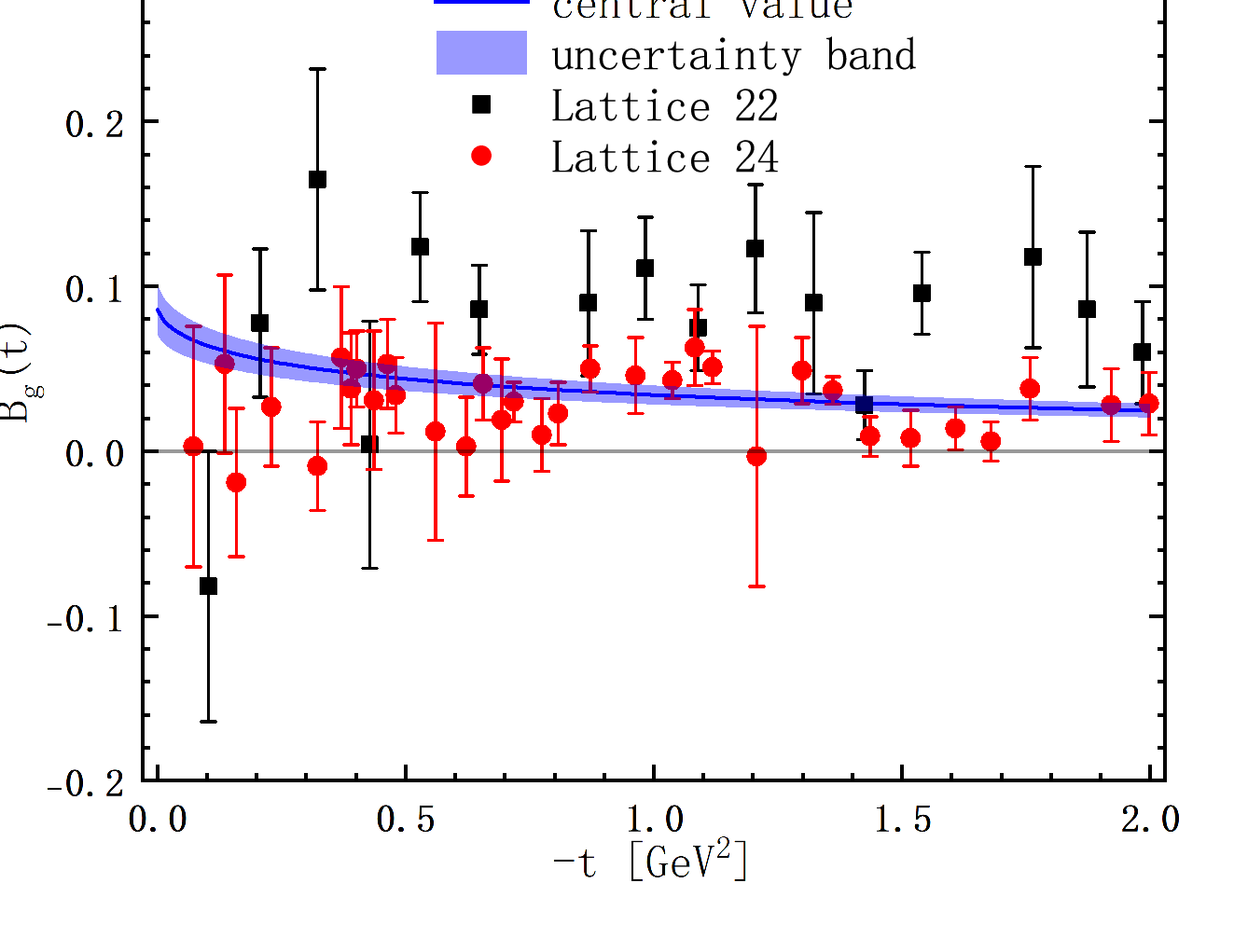}
	\\
	\includegraphics[width=0.43\columnwidth]{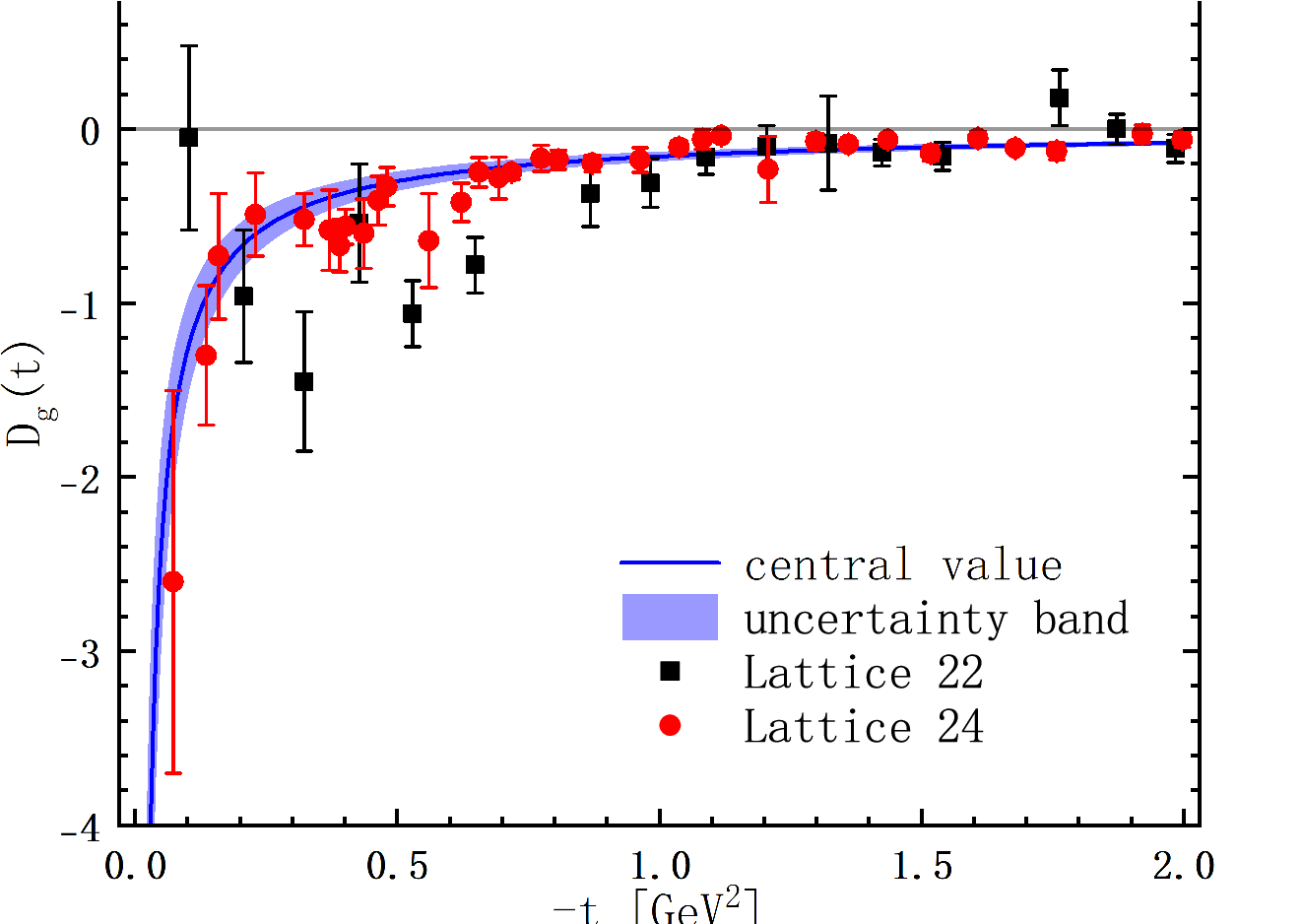}
	\includegraphics[width=0.43\columnwidth]{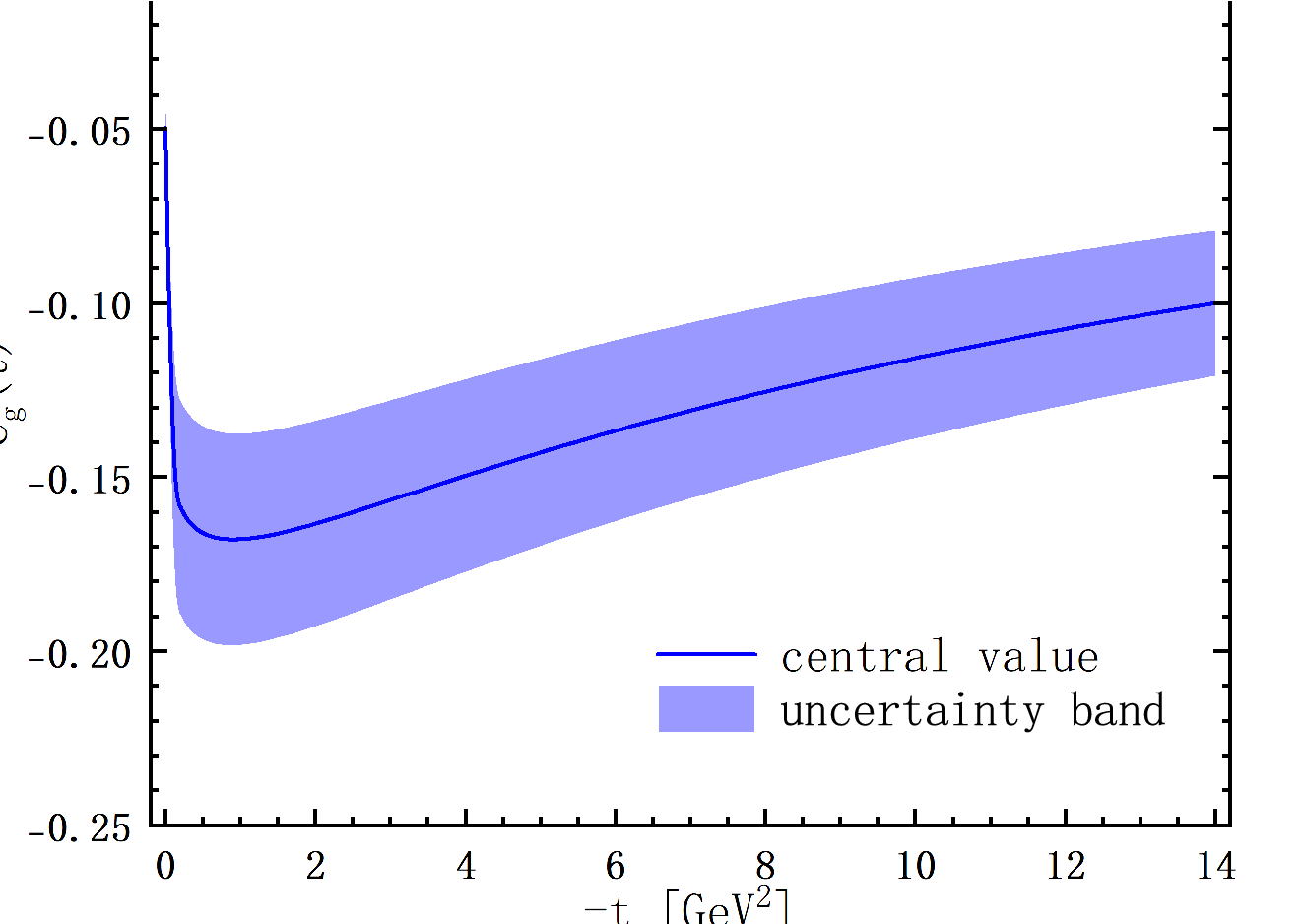}
	\\
	\caption{The gluon GFFs $A_g(t)$, $B_g(t)$, $D_g(t)$, and $\bar{C}_g(t)$ in the proton as functions of $-t$. The model results for $A_g(t)$, $B_g(t)$, and $D_g(t)$ are compared with the recent lattice QCD simulations at $\mu_0=2\,\text{GeV}$~\cite{Pefkou:2021fni,Hackett:2023rif}. The bands represent the parameter uncertainties.}
	\label{fig:GFFs}
\end{figure}

Fig.~\ref{fig:GFFs} displays the model predictions for the gluon GFFs $A_g(t)$, $B_g(t)$, $D_g(t)$, and $\bar{C}_g(t)$, where $A_g(t)$, $B_g(t)$, and $D_g(t)$ are compared with the latest lattice QCD determinations at the scale of $2\,\text{GeV}$~\cite{Pefkou:2021fni,Hackett:2023rif}.
The model prediction for $A_g(t)$ is slightly lower at small $-t$ and slightly higher at large $-t$, but is broadly consistent with the lattice results.
Our $B_g(t)$ is also in reasonable agreement with the lattice results and remains positive over the displayed range. 
By contrast, $B_g(t)$ is negative in the dressed-quark model~\cite{More:2023pcy}.
As a consequence of Poincar\'e invariance, the total GFFs $A(t)$ and $B(t)$ satisfy the sum rules $A(0)=1$ and $B(0)=0$, respectively~\cite{Lorce:2015lna,Lowdon:2017idv}.  $A(0)=1$ expresses conservation of the total momentum, while $B(0)=0$ corresponds to the vanishing of the anomalous gravitomagnetic moment for a spin-$1/2$ system~\cite{Polyakov:2018zvc}.
Comparing our gluon results with the lattice QCD determinations of the quark GFFs at the same scale~\cite{Hagler:2003jd,Gockeler:2003jfa,LHPC:2010jcs,LHPC:2007blg,
QCDSF-UKQCD:2007gdl,Deka:2013zha}, we find that the $A(t)$ sum rule is approximately satisfied within uncertainties, whereas the $B(t)$ sum rule shows a mild deviation, as listed in Table~\ref{tab2}.

\begin{table}[htbp]
	\centering
	\caption{Verification of the sum rules $A(0)=1$ and $B(0)=0$ for the total GFFs $A(t)$ and $B(t)$ based on the lattice QCD results of $A_q(0)$ and $B_q(0)$ at $\mu_0=2\,\text{GeV}$.}
	\label{tab2}
	\begin{tabular}{lccc} 
		\hline   
		$A_q(0)$ & $A_q(0)+A_g(0)(\text{this work})$ & $B_q(0)$ & $B_q(0)+B_g(0)(\text{this work})$ \\ 
		\hline 
		0.675(7)~\cite{Hagler:2003jd} &  1.068(75)  &  0   & 0.086(16)  \\
		0.547(25)~\cite{Gockeler:2003jfa} &  0.94(79)  &  0.102(137)   &  0.188(138)   \\
		0.553(43)~\cite{LHPC:2010jcs} &  0.946(86)  &  -0.077(16)   &  0.009(23)   \\
		0.520(24)~\cite{LHPC:2007blg} &  0.913(79)  &  -0.095(86)   &  -0.009(87)  \\
		0.572(12)~\cite{QCDSF-UKQCD:2007gdl} &  0.965(76)  &  -0.120(23)   &  -0.034(28)  \\
		0.565(43)~\cite{Deka:2013zha} &  0.958(86)  &  0.063(23)   &   0.149(28)  \\
		\hline 
	\end{tabular}
\end{table} 
Both $D_g(t)$ and $\bar{C}_g(t)$ are negative.
For the $D$-term form factor $D_g(t)$, our result is in good agreement with the lattice determinations.
The total GFF $D(t)$ is not associated with a Poincar\'e generator and is therefore not constrained by a corresponding sum rule; instead, it encodes the pressure and shear-force distributions inside the nucleon.
In our model, the forward limit $D_g(0)$ is divergent and tends to negative infinity.

This behavior differs qualitatively from that of commonly used multipole parametrizations, which yield finite negative values from fits to lattice QCD results~\cite{Pefkou:2021fni,Hackett:2023rif,Shanahan:2018pib} and phenomenological extractions~\cite{Duran:2022xag,Guo:2021ibg,Guo:2023pqw,Wang:2022ndz}.
However, a similar divergent behavior appears in Ref.~\cite{More:2023pcy}, even though $D_g(t)$ in the dressed quark model is a positive distribution.
Note that the GFF $D_\gamma(t)$ for the photon part of the EMT in quantum electrodynamics was also found to be divergent in a Feynman diagram approach~\cite{Freese:2022jlu,Metz:2021lqv}, which was attributed to the infinite range of the Coulomb interaction~\cite{Berends:1975ah,Milton:1976jr}.
Because $D_g(0)$ is divergent and the slope of $A_g(t)$ at $t=0$ is larger than suggested by the lattice data~\cite{Pefkou:2021fni,Hackett:2023rif}, our model does not yield reliable numerical estimates for gluonic 3D or 2D radii, including the mechanical, mass, and scalar radii.
At present, no lattice QCD or experimental determinations of $\bar{C}_g(t)$ are available.

Although the total $\bar{C}(t)$ must identically equal zero, the individual contribution can be a small nonzero value that depends on the scale, as discussed for the small value of $\bar{C}_q(0)$ using the instanton picture of the QCD vacuum~\cite{Polyakov:2018exb}, which might justify Teryaev's equipartition conjecture~\cite{Teryaev:2006fk,Teryaev:2016edw}.
In our model, the gluon GFF $\bar{C}_g(t)$ has a negative peak at small $-t$ and converges to zero much more slowly than the other GFFs as $-t$ increases. 
In the forward limit, we find that $\bar{C}_g(0)=-0.050(9)$, the sign of which is the same as that in other models at $\mu_0=2\,\text{GeV}$, specifically, $\bar{C}_g(0)=-0.0146$ in the dressed quark model~\cite{More:2023pcy}, and $\bar{C}_g(0)=-0.030(10)$ in the light-front spectator model motivated by soft-wall AdS/QCD~\cite{Sain:2025kup}.
Meanwhile, Ref.~\cite{More:2023pcy} reported a value of $\bar{C}_q(0)=0.0146$, and Ref.~\cite{Polyakov:2018exb} reported a value of $\bar{C}_q(0)=0.014$ at the scale of $0.4\,\text{GeV}^2$.

\begin{figure}
	\centering
	\includegraphics[width=0.45\columnwidth]{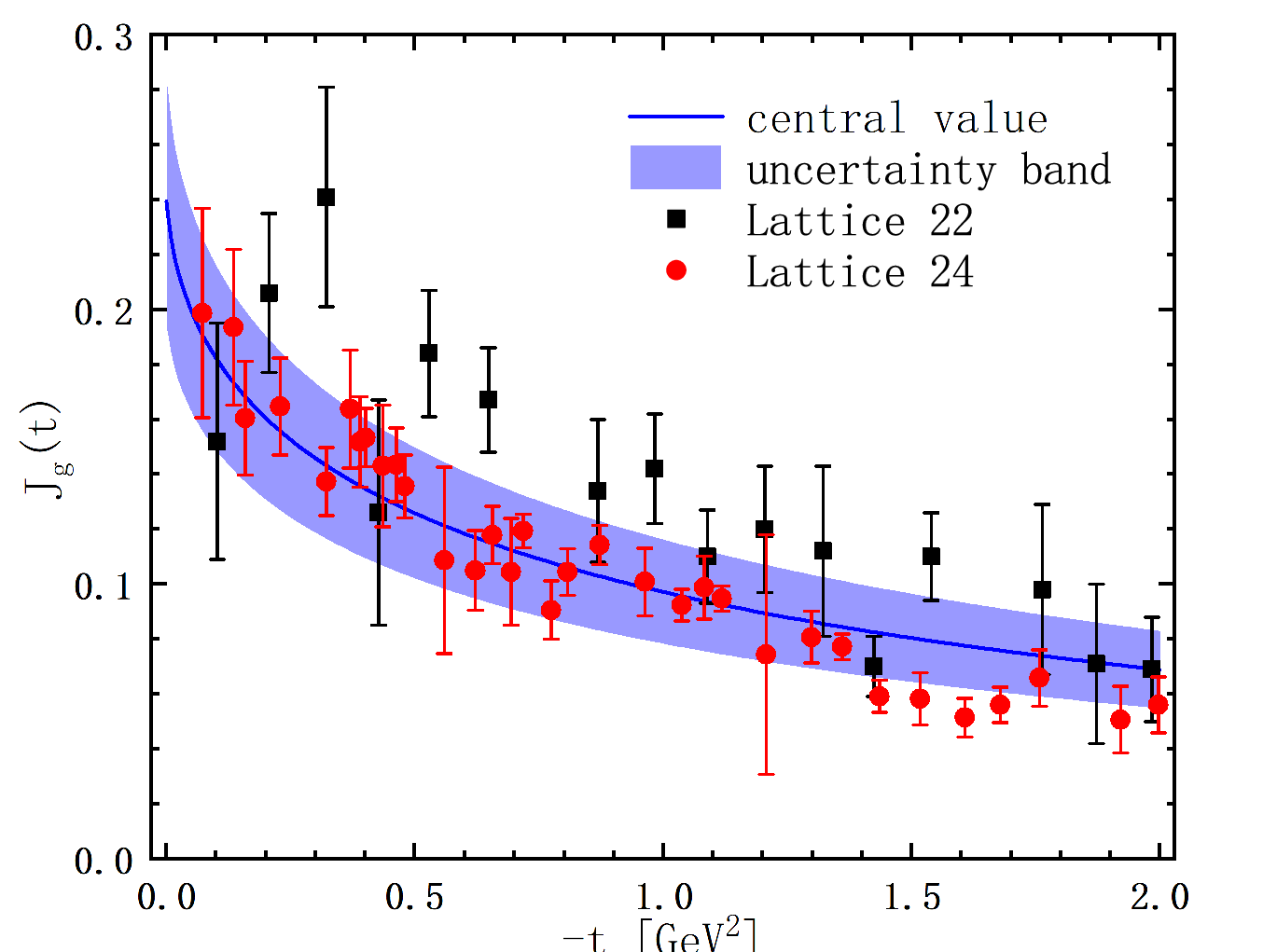}
	\caption{The gluon angular momentum GFF $J_g(t)$ in the proton as a function of $-t$ compared with the recent lattice QCD simulations at $\mu_0=2\,\text{GeV}$~\cite{Pefkou:2021fni,Hackett:2023rif}. The band represents the parameter uncertainties.}
	\label{fig:Jgt}
\end{figure}

According to Ji's sum rule~\cite{Ji:1996ek}, the contribution to the proton's total angular momentum from gluons/quarks can be obtained by
\begin{align}
	J_i(t=0) = \frac{1}{2}\left[A_i(0) + B_i(0)\right]\label{eq:Ji}.
\end{align}

Fig.~\ref{fig:Jgt} shows the gluon angular-momentum GFF $J_g(t)$ as a function of $-t$. The solid curve and band correspond to the central parameter values and their uncertainties, respectively. 
Our model is in reasonable agreement with the latest lattice QCD results~\cite{Pefkou:2021fni,Hackett:2023rif}.
As $-t$ increases, $J_g(t)$ smoothly approaches zero while remaining positive.
By contrast, the same GFF exhibits a zero crossing in the large $-t$ region in Ref.~\cite{Sain:2025kup}. 
The gluon angular momentum obtained from Eq.~(\ref{eq:Ji}) in our model is $J_g=0.239(45)$, consistent within uncertainties with recent lattice QCD determinations at $\mu_0=2\,\text{GeV}$: $J_g=0.231(11)(22)$~\cite{Wang:2021vqy}, $J_g=0.187(46)$~\cite{Alexandrou:2020sml}, and $J_g=0.255(13)$~\cite{Hackett:2023rif}.
Additionally, the Bethe-Salpeter approach also provides a similar result at the same scale: $J_g=0.208(6)$~\cite{Yao:2024ixu}.

\begin{figure}
	\centering
	\includegraphics[width=0.45\columnwidth]{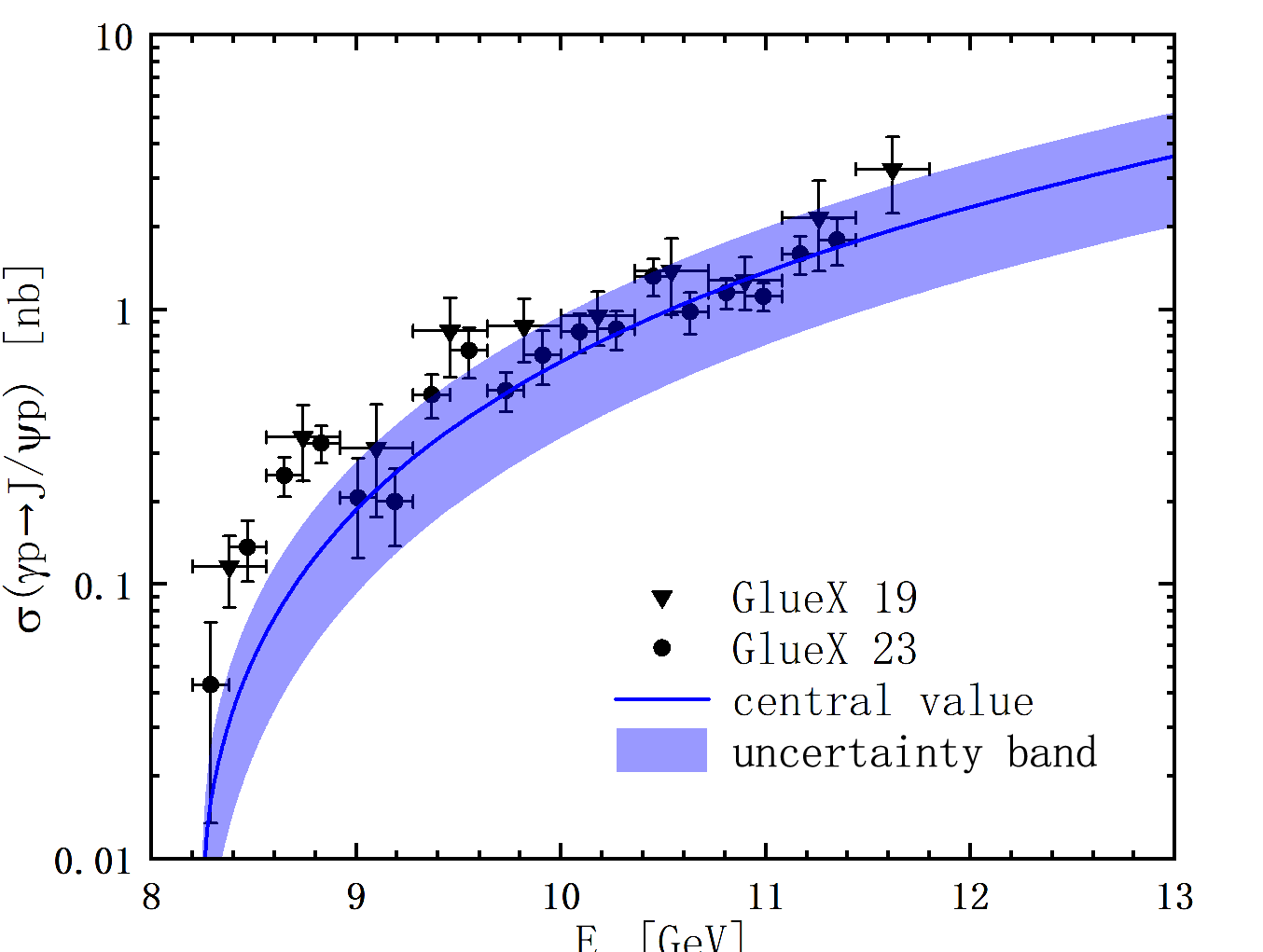}
	\caption{The total cross section for $\gamma p \rightarrow J/\psi\,p$ as a function of $E_\gamma$ compared with data from GlueX19~\cite{GlueX:2019mkq} (triangles) and GlueX23~\cite{GlueX:2023pev} (circles). The band represents the parameter uncertainties.}
	\label{fig:tot}
\end{figure}

To further test the parameter set in Table~\ref{tab1}, we compare the model prediction  for the total cross section in Eq.~(\ref{eq:sigma}) with the experimental measurements~\cite{GlueX:2019mkq,GlueX:2023pev} in Fig.~\ref{fig:tot}.
The model reproduces the overall magnitude and energy dependence of the data, with some deviation for $E_\gamma<9\,\text{GeV}$, where differential cross section data are unavailable. 
This provides an additional check of the fitted parameter set.

\subsection{Light-front densities}

The GFFs $A_i(q^2)$ and $D_i(q^2)$ encode a range of mechanical properties of the hadron.
The 2D light-front densities in the impact parameter space $\bm{r}_\perp$--namely the light-front momentum density $\varepsilon_i^{(\text{LF})}(r_\perp)$, the angular momentum density $\rho_{J,i}^{(\text{LF})}(r_\perp)$, the pressure density $p_i^{(\text{LF})}(r_\perp)$, and the shear force density $s_i^{(\text{LF})}(r_\perp)$--are defined as~\cite{Freese:2021czn,Kim:2021jjf}
\begin{align}
	\varepsilon_i^{(\text{LF})}(r_\perp) &= P^+ \tilde{A}_i(r_\perp),\label{eq:epsilon}\\
	\rho_{J,i}^{(\text{LF})}(r_\perp) &= -\frac{1}{2} r_\perp \frac{\mathrm{d}}{\mathrm{d}r_\perp} \tilde{J}_i(r_\perp),\\
	p_i^{(\text{LF})}(r_\perp) &= \frac{1}{8P^+} \frac{1}{r_\perp} \frac{\mathrm{d}}{\mathrm{d}r_\perp} r_\perp \frac{\mathrm{d}}{\mathrm{d}r_\perp} \tilde{D}_i(r_\perp),\label{eq:pi}
\end{align}
\begin{align}	
	s_i^{(\text{LF})}(r_\perp) = -\frac{1}{4P^+} r_\perp \frac{\mathrm{d}}{\mathrm{d}r_\perp} \frac{1}{r_\perp} \frac{\mathrm{d}}{\mathrm{d}r_\perp} \tilde{D}_i(r_\perp),\label{eq:si}
\end{align}	
where Eqs.~(\ref{eq:pi}-\ref{eq:si}) show that the pressure and shear force distributions are not independent, and $\tilde{F}_i(r_\perp)$ are obtained by performing a 2D Fourier transform of the GFFs $F_i(-\bm{q}_\perp^2)$:
\begin{align}
	\tilde{F}_i(r_\perp) = \int \frac{\mathrm{d}^2 \bm{q}_\perp}{(2\pi)^2} e^{-i\bm{q}_\perp\cdot\bm{r}_\perp} F_i(-\bm{q}_\perp^2).
\end{align}	
Here, the 2D momentum and position vectors $\bm{q}_\perp$ and $\bm{r}_\perp$ are transverse to the direction of motion of the hadron.
For convenience, the expressions in Eqs.~(\ref{eq:epsilon}) and (\ref{eq:pi}-\ref{eq:si}) are globally rescaled by multiplication with the Lorentz factors~\cite{Panteleeva:2021iip}:
\begin{align}
	\mathcal{E}_i(r_\perp) &= \frac{M}{P^+} \varepsilon_i^{(\text{LF})}(r_\perp),\label{eq:E}\\
	\mathcal{P}_i(r_\perp) &= \frac{P^+}{2M} p_i^{(\text{LF})}(r_\perp), \\
	\mathcal{S}_i(r_\perp) &= \frac{P^+}{2M} s_i^{(\text{LF})}(r_\perp),
\end{align}	
where $\mathcal{E}_i(r_\perp)$ is the 2D light-front mass density and describes the transverse distribution of the parton mass. 
The Abel transformation relates the 3D and 2D distributions while preserving their physical interpretation. 
Because a longitudinal boost does not mix the longitudinal component of angular momentum, the corresponding density requires no rescaling~\cite{Lorce:2017wkb}.
After integrating $\mathcal{E}_i(r_\perp)$ and $\rho_{J,i}^{(\text{LF})}(r_\perp)$ over $\bm{r}_\perp$, one obtains
\begin{align}
	\int d^2\bm{r}_\perp\mathcal{E}_i(r_\perp)=MA_i(0),\qquad \int d^2\bm{r}_\perp \rho_{J,i}^{(\text{LF})}(r_\perp)=J_i(0),
\end{align}	
which, after summing over all constituents, lead to the sum rules $A(0)=A_q(0)+A_g(0)=1$ and $J(0)=J_q(0)+J_g(0)=1/2$, respectively, with the latter being exactly the spin of the nucleon.
Additionally, the radial and tangential forces of the parton within a hadron are defined as~\cite{Lorce:2018egm,Freese:2021qtb}
\begin{align}
	\mathcal{P}_{r,i}(r_\perp) &= \mathcal{P}_i(r_\perp) + \frac{1}{2} \mathcal{S}_i(r_\perp), \\
	\mathcal{P}_{t,i}(r_\perp) &= \mathcal{P}_i(r_\perp) - \frac{1}{2} \mathcal{S}_i(r_\perp),
\end{align}	
where $\mathcal{P}_{r,i}(r_\perp)$ is called the ``radial force" to emphasize the absence of a net force across the hadron~\cite{Sain:2025kup}.

\begin{figure}
	\centering
	\includegraphics[width=0.43\columnwidth]{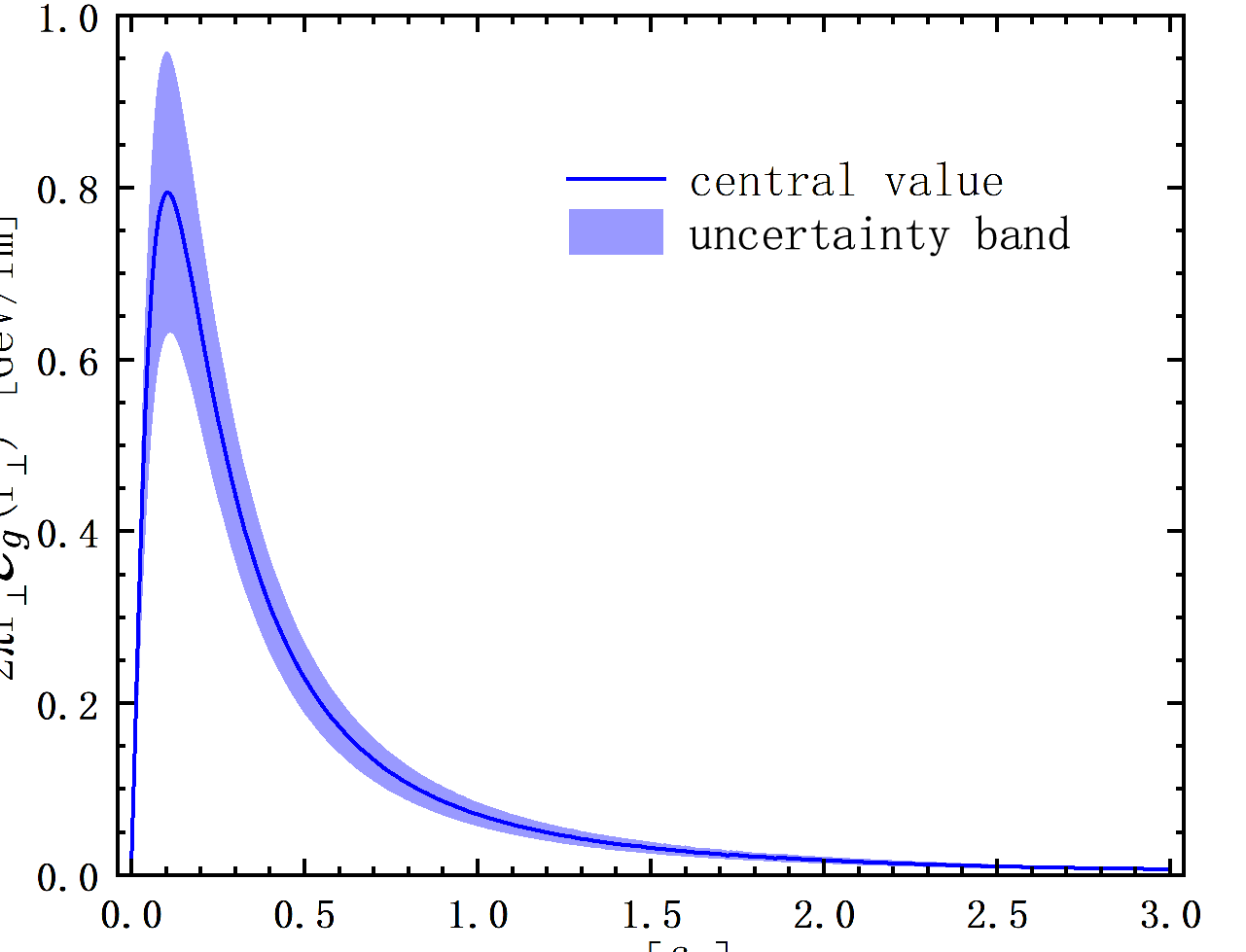}
	\includegraphics[width=0.43\columnwidth]{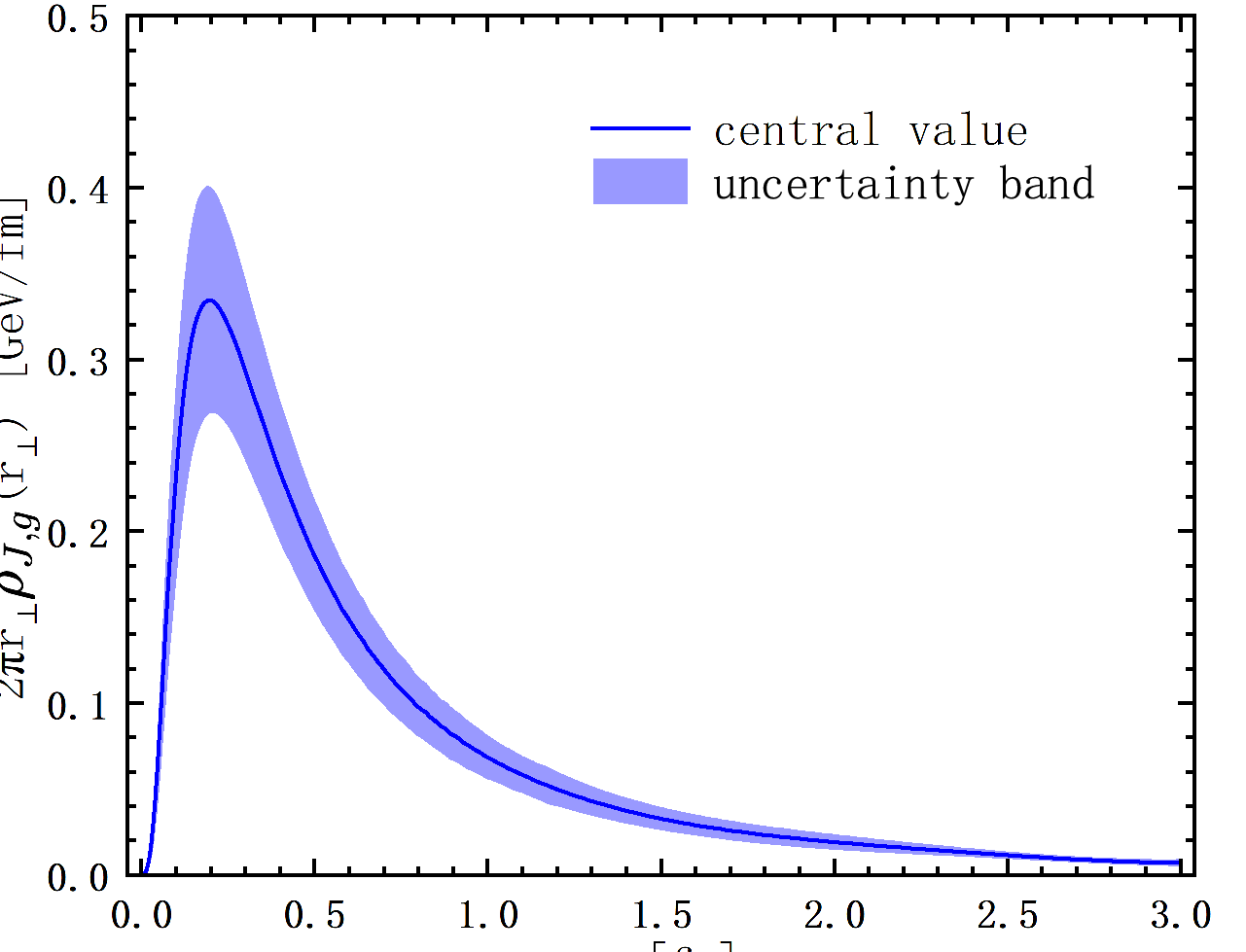}
	\\
	\caption{The 2D light-front gluonic mass $\mathcal{E}_g(r_\perp)$ (left panel) and angular momentum $\mathcal{\rho}_{J,g}(r_\perp)$ (right panel) distributions weighted by $2\pi r_\perp$ as functions of $r_\perp$. The bands represent the parameter uncertainties.}
	\label{fig:Erho}
\end{figure}

\begin{figure}
	\centering
	\includegraphics[width=0.43\columnwidth]{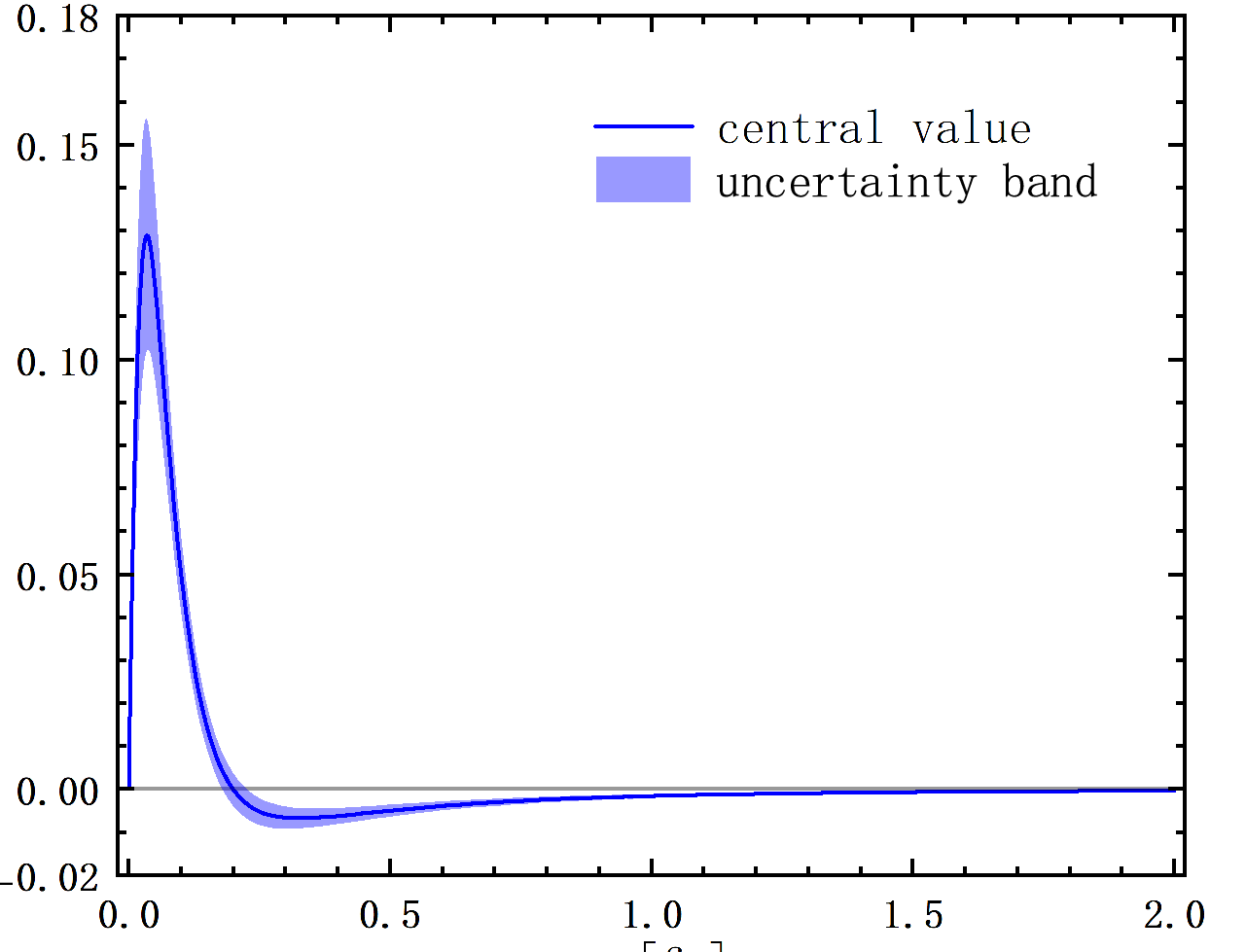}
	\includegraphics[width=0.43\columnwidth]{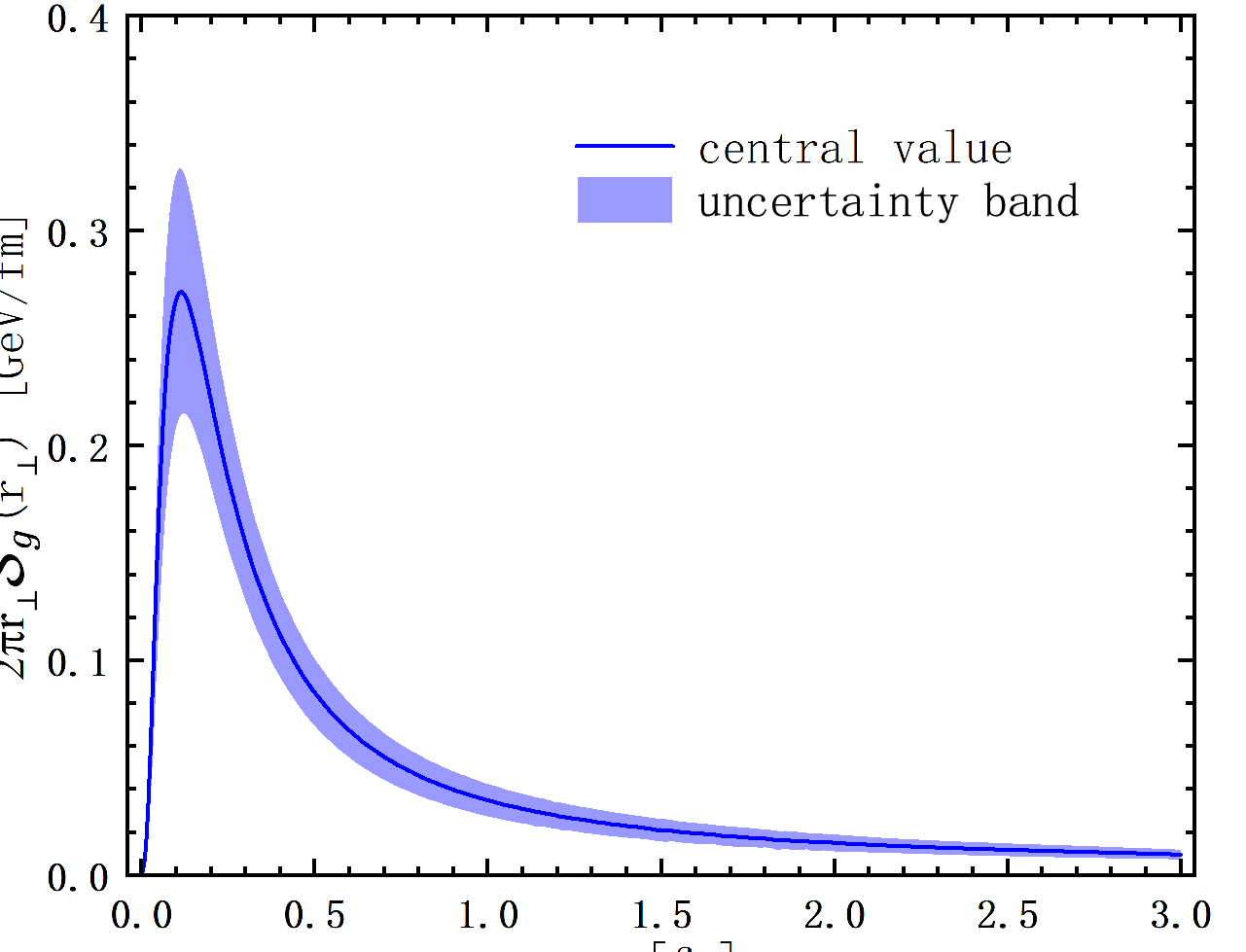}
	\\
	\caption{The 2D light-front gluonic pressure $\mathcal{P}_g(r_\perp)$ (left panel) and shear force $\mathcal{S}_g(r_\perp)$ (right panel) distributions weighted by $2\pi r_\perp$ as functions of $r_\perp$. The bands represent the parameter uncertainties.}
	\label{fig:PS}
\end{figure}

\begin{figure}
	\centering
	\includegraphics[width=0.43\columnwidth]{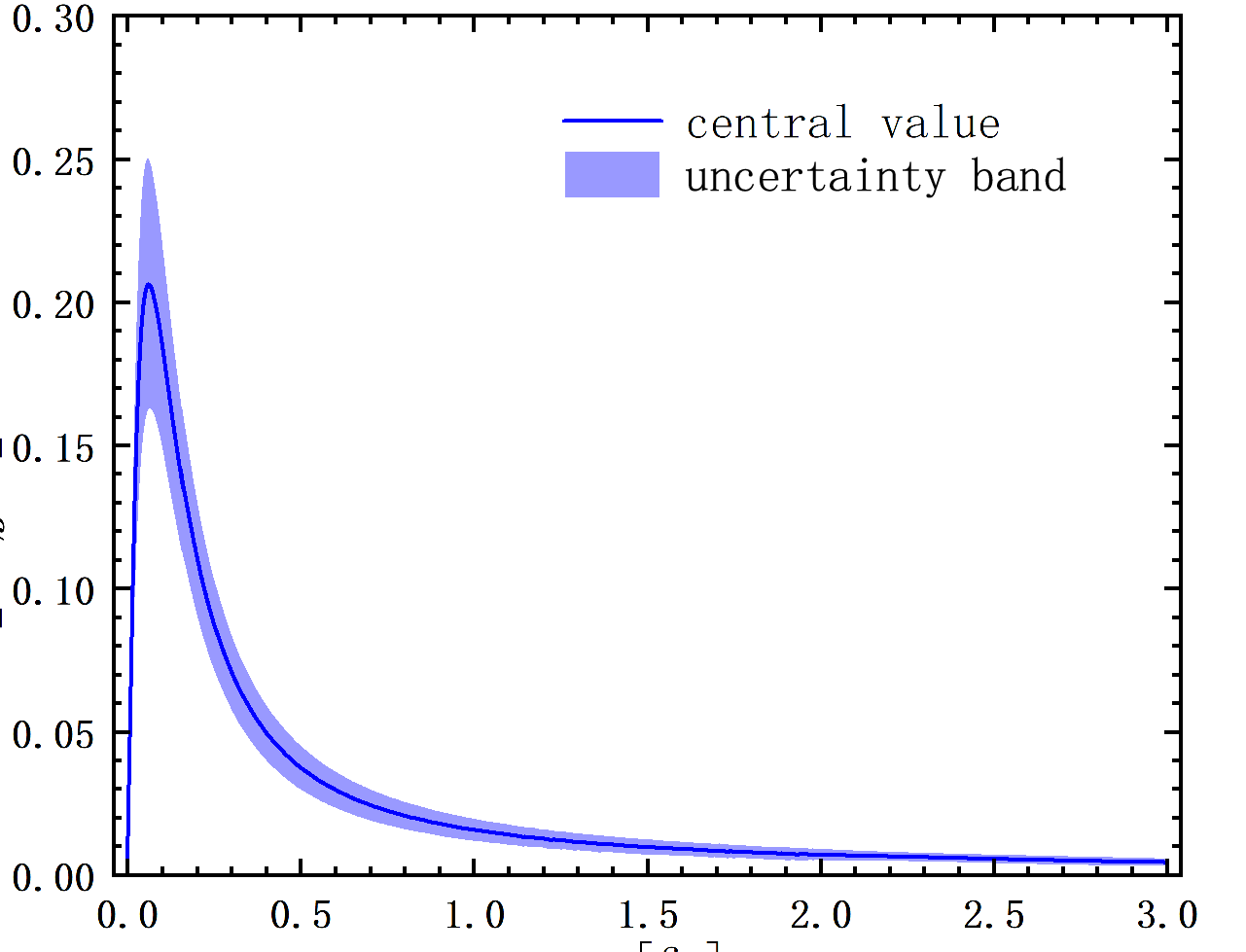}
	\includegraphics[width=0.43\columnwidth]{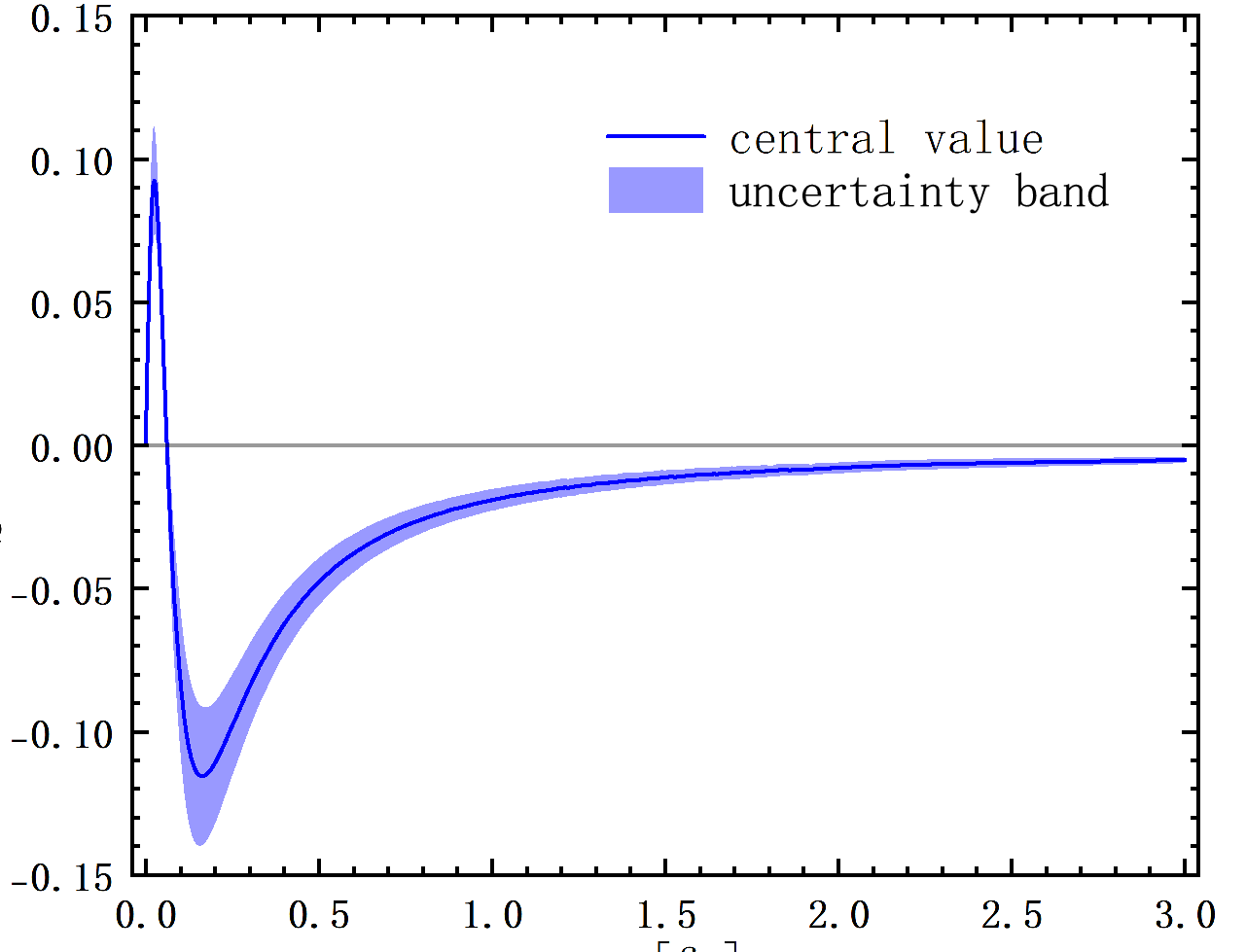}
	\\
	\caption{The 2D light-front gluonic radial force $\mathcal{P}_{r,g}(r_\perp)$ (left panel) and tangential force $\mathcal{P}_{t,g}(r_\perp)$ (right panel) distributions weighted by $2\pi r_\perp$ as functions of $r_\perp$. The bands represent the parameter uncertainties.}
	\label{fig:PrPt}
\end{figure}

Figs.~\ref{fig:Erho}-\ref{fig:PrPt} show the 2D light-front gluonic mass $\mathcal{E}_g(r_\perp)$, angular momentum $\rho_{J,g}(r_\perp)$, pressure $\mathcal{P}_g(r_\perp)$, shear force $\mathcal{S}_g(r_\perp)$, radial force $\mathcal{P}_{r,g}(r_\perp)$, and tangential force $\mathcal{P}_{t,g}(r_\perp)$ distributions in the proton, 
each weighted by $2\pi r_\perp$ and thus representing the contribution from a circle of radius $r_\perp$.
The dominant contributions are concentrated near the proton core.
They vanish at the center, $r_\perp=0$, and decrease again toward zero at large transverse distances. 
Compared with the light-front spectator model motivated by soft-wall AdS/QCD~\cite{Sain:2025kup} and with the tripole and $z$-expansion parametrizations~\cite{Pefkou:2021fni}, the distributions from our model exhibit similar qualitative behavior with a narrower profile.

In Fig.~\ref{fig:Erho}, the light-front gluon mass distribution (left panel) peaks around $r_\perp=0.1\,\text{fm}$. At the proton center, we obtain $\mathcal{E}_g(0)=2.446\pm 0.678\,\text{GeV}/\text{fm}^2$, which is larger than that of the quark $\mathcal{E}_q(0)=1.61\,\text{GeV}/\text{fm}^2$~\cite{Kim:2021jjf}.
The light-front gluon angular momentum distribution (right panel) remains positive over the entire $r_\perp$ region due to the positive definiteness of $J_g(t)$ shown in Fig.~\ref{fig:Jgt}.
In contrast, $\rho_{J,g}(r_\perp)$ in Ref.~\cite{Sain:2025kup} is negative near the center and then becomes positive at larger transverse distances, which arises from the negative distribution of $J_g(t)$ at large $-t$. 
Quantitatively, based on the model results for the light-front quark distributions~\cite{Kim:2021jjf}, both the mass and angular momentum densities are overall lower for gluons than for quarks, especially the mass density.

In Fig.~\ref{fig:PS}, the light-front gluon pressure distribution (left panel) is positive in the region near $r_\perp=0\,\text{fm}$, peaking around $r_\perp=0.03\,\text{fm}$, and then becomes negative at larger transverse distances, peaking around $r_\perp=0.3\,\text{fm}$, with the nodal point located at $r_0\sim0.2\,\text{fm}$, indicating that the vector field reverses direction at this point.
This behavior complies with the qualitative requirement of the 2D von Laue stability condition~\cite{Polyakov:2018zvc,Freese:2021czn},
\begin{align}
	\int \mathrm{d}^2\bm{r}_\perp \mathcal{P}_i(r_\perp) &=0
	\label{eq:von},
\end{align}
which ensures the balance of internal forces in a stable system. In our model, however, a precise numerical verification is hindered by sizable oscillations at very large $r_\perp$.
The similar behavior of the quark pressure distribution~\cite{Burkert:2018bqq,Kim:2021jjf,Nair:2024fit} suggests a picture of mechanical balance in which the repulsive inner region is compensated by the confining pressure in the outer region.
In the 3D Breit frame, the gluon pressure distribution always has a wider positive peak and a larger nodal point~\cite{Shanahan:2018nnv,Yao:2024ixu,Pefkou:2021fni}.
At the proton center, the pressure is given by
\begin{align}
	\mathcal{P}_g(0) = \int_{0}^{\infty} \frac{dr_{\perp}}{r_{\perp}} \mathcal{S}_g(r_{\perp})=1.210\pm 0.344\,\frac{\text{GeV}}{\text{fm}^2},
\end{align}
which is significantly larger than that of the quark:  $\mathcal{P}_q(0)=0.045\,\text{GeV}/\text{fm}^2$~\cite{Kim:2021jjf}.
The light-front gluon shear force (right panel) shows a positive distribution over the entire $r_\perp$ region and peaks around $r_\perp=0.1\,\text{fm}$, consistent with the 2D local stability condition $\mathcal{S}_g(r_\perp)>0$~\cite{Panteleeva:2021iip}.
However, the 3D shear force distribution in the Breit frame is negative near the center up to $r_\perp\approx 0.5\,\text{fm}$~\cite{Shanahan:2018nnv}.
The shear force on the light front is related to the 2D line tension (the 2D counterpart of the surface tension in 3D)~\cite{Panteleeva:2021iip}:
\begin{align}	\gamma^{(\text{2D})}=\int_{0}^{\infty}dr_{\perp}\mathcal{S}_g(r_{\perp})=0.102\pm0.022\,\frac{\text{GeV}}{\text{fm}}.
\end{align}	
A positive value of this quantity corresponds to the constraint on the surface tension for stable hydrostatic systems in 3D~\cite{Polyakov:2018zvc}.
Additionally, the 2D equilibrium condition from the conservation of the gluon EMT is also satisfied~\cite{Freese:2021czn}:
\begin{align}	
	\frac{\mathrm{d}\mathcal{P}_i(r_\perp)}{\mathrm{d}r_\perp} + \frac{1}{2}\frac{\mathrm{d}\mathcal{S}_i(r_\perp)}{\mathrm{d}r_\perp} + \frac{1}{r_\perp} \mathcal{S}_i(r_\perp) = 0,\label{eq:PSd}
\end{align}
which does not depend on the specific forms of $\mathcal{P}_i(r_\perp)$ and $\mathcal{S}_i(r_\perp)$.

In Fig.~\ref{fig:PrPt}, the light-front gluon radial force distribution (left panel) remains positive over the entire $r_\perp$ region.
This result is in accordance with the 2D local stability condition $\mathcal{P}_{r,i}(r_\perp)\geq0$~\cite{Lorce:2018egm,Freese:2021czn},
which is nontrivial and implies the Polyakov-Schweitzer negativity condition~\cite{Polyakov:2018zvc}, i.e., $D_i(0)\leq0$.
Similar to the pressure distribution, the light-front gluon tangential force (right panel) is positive in the central region, corresponding to a repulsive force, and negative in the outer region, corresponding to an attractive force. The positive and negative regions peak around $r_\perp=0.02\,\text{fm}$ and $0.16\,\text{fm}$, respectively, with the latter having a larger magnitude; the nodal point is at $r_0\sim0.06\,\text{fm}$.
This distribution can be numerically verified to comply with the von Laue stability condition in the 1D subsystem~\cite{Kim:2021jjf}:
\begin{align}	
	\int_0^\infty dr_\perp \mathcal{P}_{t,i}(r_\perp)=0
	\label{eq:von1},
\end{align}
which helps maintain stability in the tangential direction of the proton.
Given the similar shapes of the light-front quark radial and tangential force distributions~\cite{Kim:2021jjf,Nair:2024fit}, the total strong force field near the proton's core is dominated by the radial force, whereas the outer region is dominated by the tangential force, which ensures that the proton is stable.
Note that the dominance of the 3D tangential force in the Breit frame at large distances can be obtained model-independently~\cite{Alharazin:2020yjv}.

\subsection{Galilean energy density and pressure distributions}

The 2D Galilean energy density, radial pressure, tangential pressure, isotropic pressure, and pressure anisotropy for quarks/gluons in the proton are defined as~\cite{Lorce:2018egm}
\begin{align}
	\mu_i(r_\perp) &= M\left[\frac{\tilde{A}_i(r_\perp)}{2} + \tilde{\bar{C}}_i(r_\perp) + \frac{1}{4M^2}\frac{1}{r_\perp}\frac{\mathrm{d}}{\mathrm{d}r_\perp}\left(r_\perp\frac{\mathrm{d}}{\mathrm{d}r_\perp}\left[\frac{\tilde{B}_i(r_\perp)}{2} - 4\tilde{C}_i(r_\perp)\right]\right)\right],\label{eq:mui}\\
	\sigma_{r,i}(r_\perp) &= M\left[-\tilde{\bar{C}}_i(r_\perp) + \frac{1}{M^2}\frac{1}{r_\perp}\frac{\mathrm{d}\tilde{C}_i(r_\perp)}{\mathrm{d}r_\perp}\right],\\
	\sigma_{t,i}(r_\perp) &= M\left[-\tilde{\bar{C}}_i(r_\perp) + \frac{1}{M^2}\frac{\mathrm{d}^2\tilde{C}_i(r_\perp)}{\mathrm{d}r_\perp^2}\right],
\end{align}
\begin{align}		
	\sigma_i(r_\perp) &= M\left[-\tilde{\bar{C}}_i(r_\perp) + \frac{1}{2M^2}\frac{1}{r_\perp}\frac{\mathrm{d}}{\mathrm{d}r_\perp}\left(r_\perp\frac{\mathrm{d}\tilde{C}_i(r_\perp)}{\mathrm{d}r_\perp}\right)\right],\label{eq:sigmai}\\
	\Pi_i(r_\perp) &= M\left[-\frac{1}{M^2}r_\perp\frac{\mathrm{d}}{\mathrm{d}r_\perp}\left(\frac{1}{r_\perp}\frac{\mathrm{d}\tilde{C}_i(r_\perp)}{\mathrm{d}r_\perp}\right)\right]\label{eq:Pii},
\end{align}
where the Galilean energy density includes contributions from all four parton GFFs, and the isotropic pressure and the pressure anisotropy can also be defined in terms of the radial and tangential pressures as follows:
\begin{align}
	\sigma_i(r_\perp) &= \frac{\sigma_{r,i}(r_\perp) + \sigma_{t,i}(r_\perp)}{2}, \label{eq:sigmairt}\\
	\Pi_i(r_\perp) &= \sigma_{r,i}(r_\perp) - \sigma_{t,i}(r_\perp)\label{eq:Piirt}.
\end{align}

\begin{figure}
	\centering
	\includegraphics[width=0.45\columnwidth]{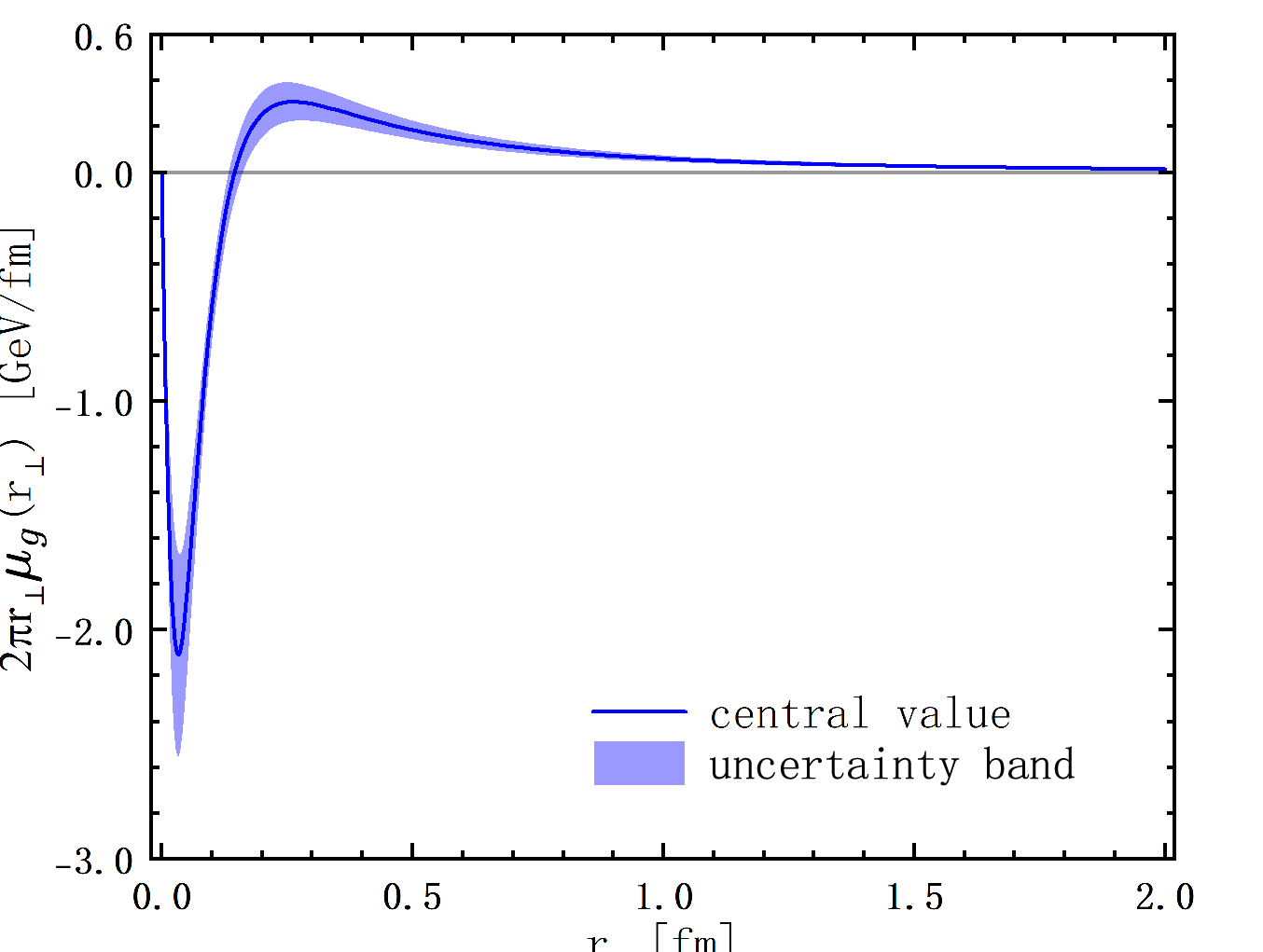}
	\caption{The 2D gluonic Galilean energy distribution $\mu_g(r_\perp)$ weighted by $2\pi r_\perp$ as a function of $r_\perp$. The band represents the parameter uncertainties.}
	\label{fig:mu}
\end{figure}

\begin{figure}
	\centering
	\includegraphics[width=0.43\columnwidth]{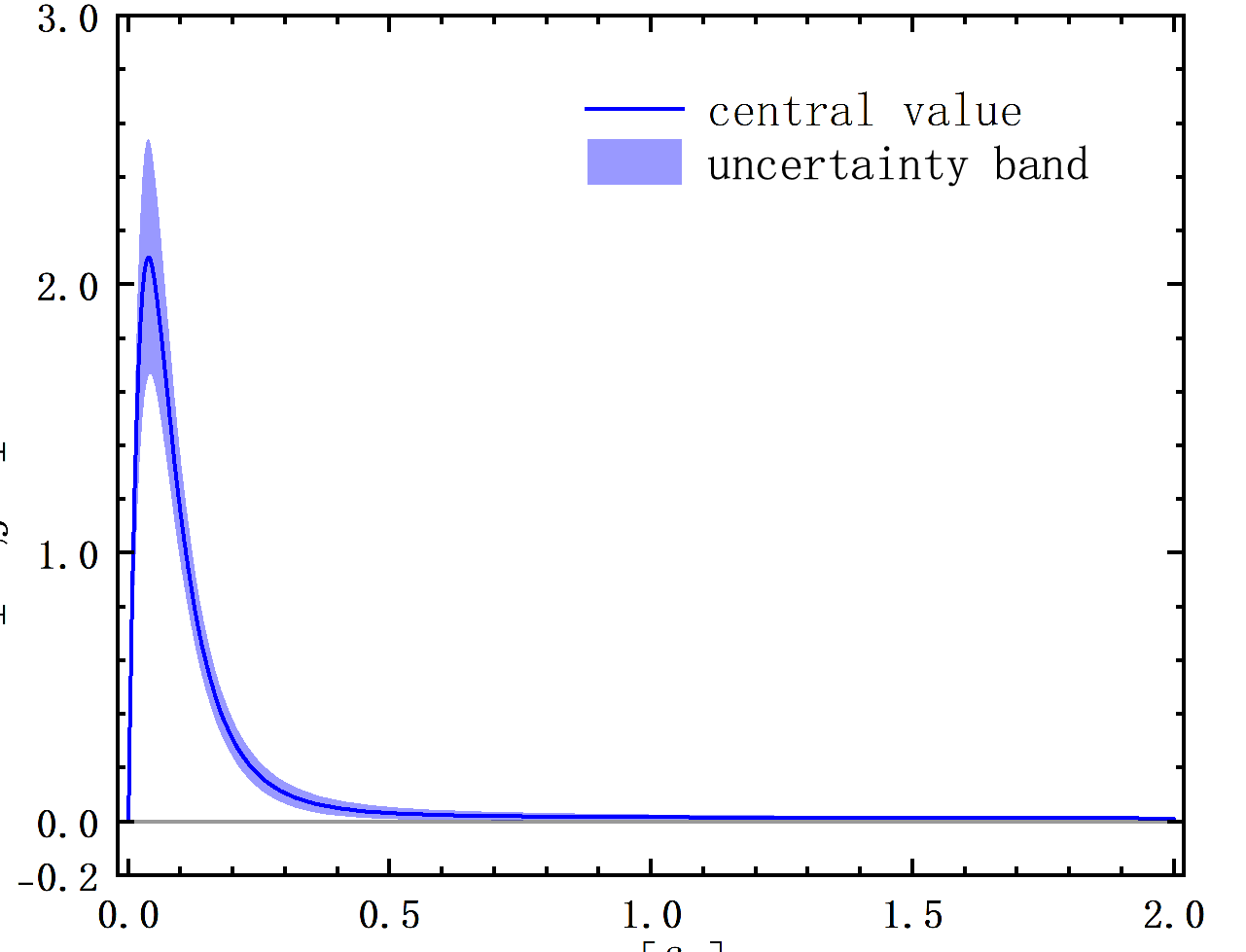}
	\includegraphics[width=0.43\columnwidth]{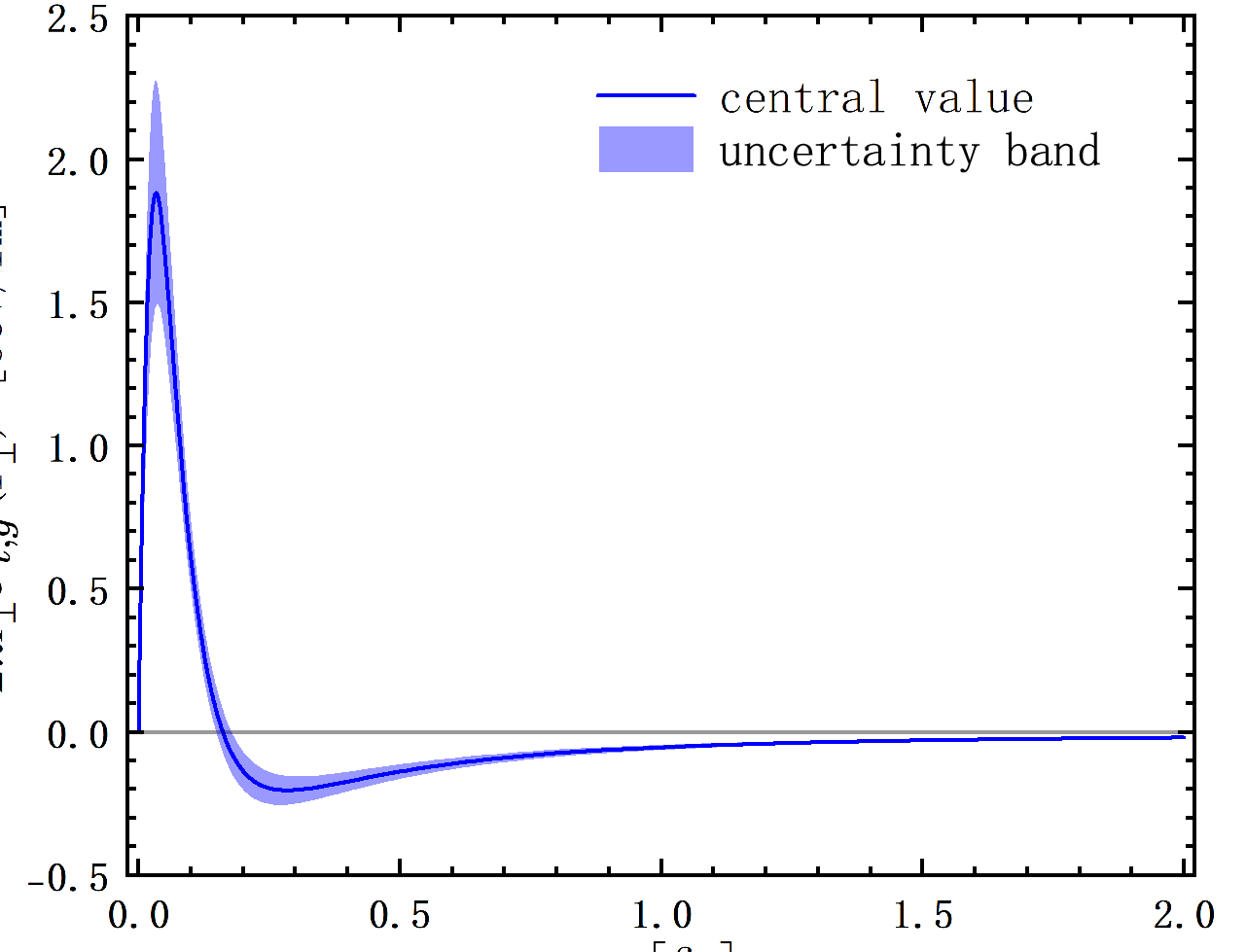}
	\\
	\caption{The 2D gluonic radial pressure $\sigma_{r,g}(r_\perp)$ (left panel) and tangential pressure $\sigma_{t,g}(r_\perp)$ (right panel) distributions weighted by $2\pi r_\perp$ as functions of $r_\perp$. The bands represent the parameter uncertainties.}
	\label{fig:sigmart}
\end{figure}

\begin{figure}
	\centering
	\includegraphics[width=0.43\columnwidth]{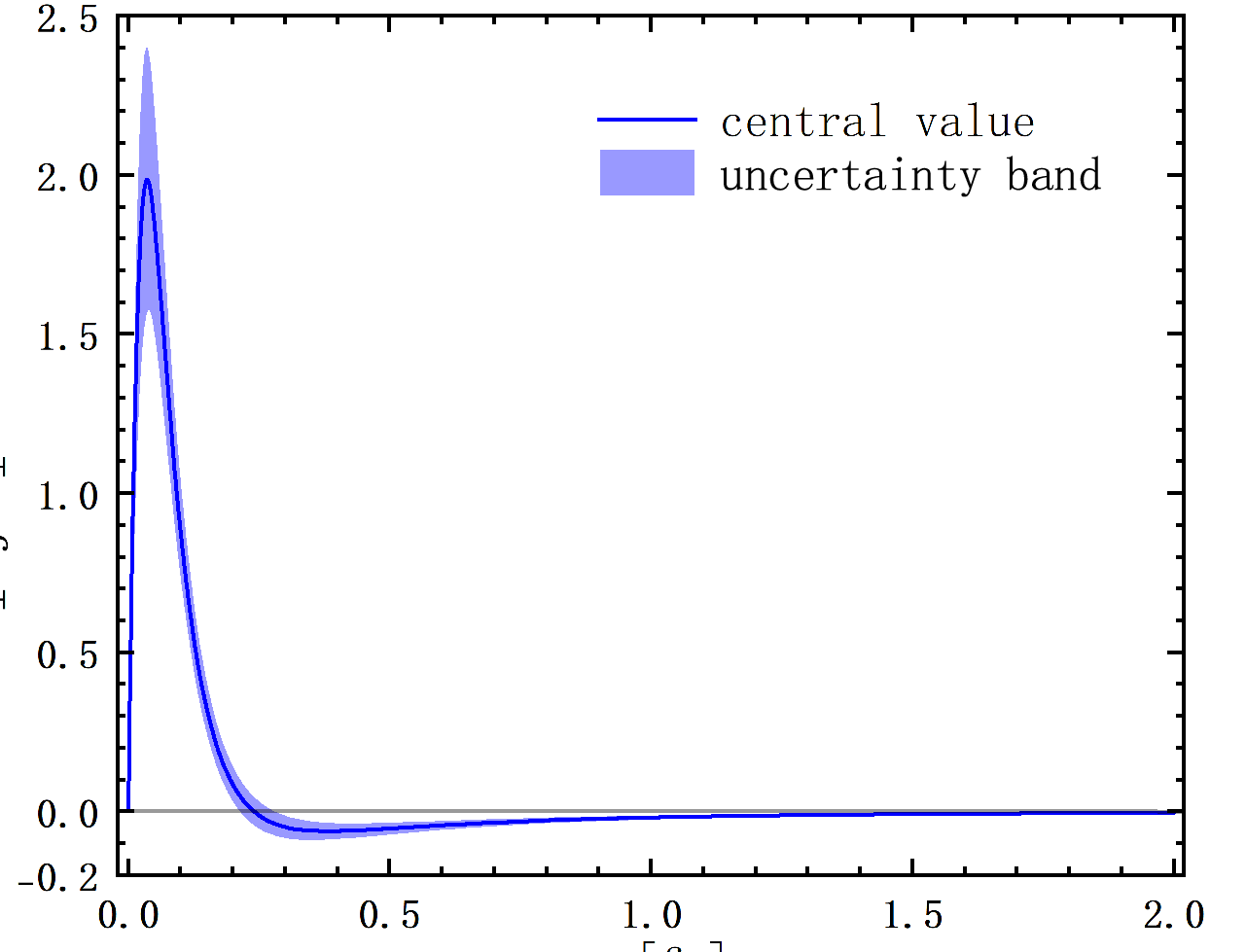}
	\includegraphics[width=0.43\columnwidth]{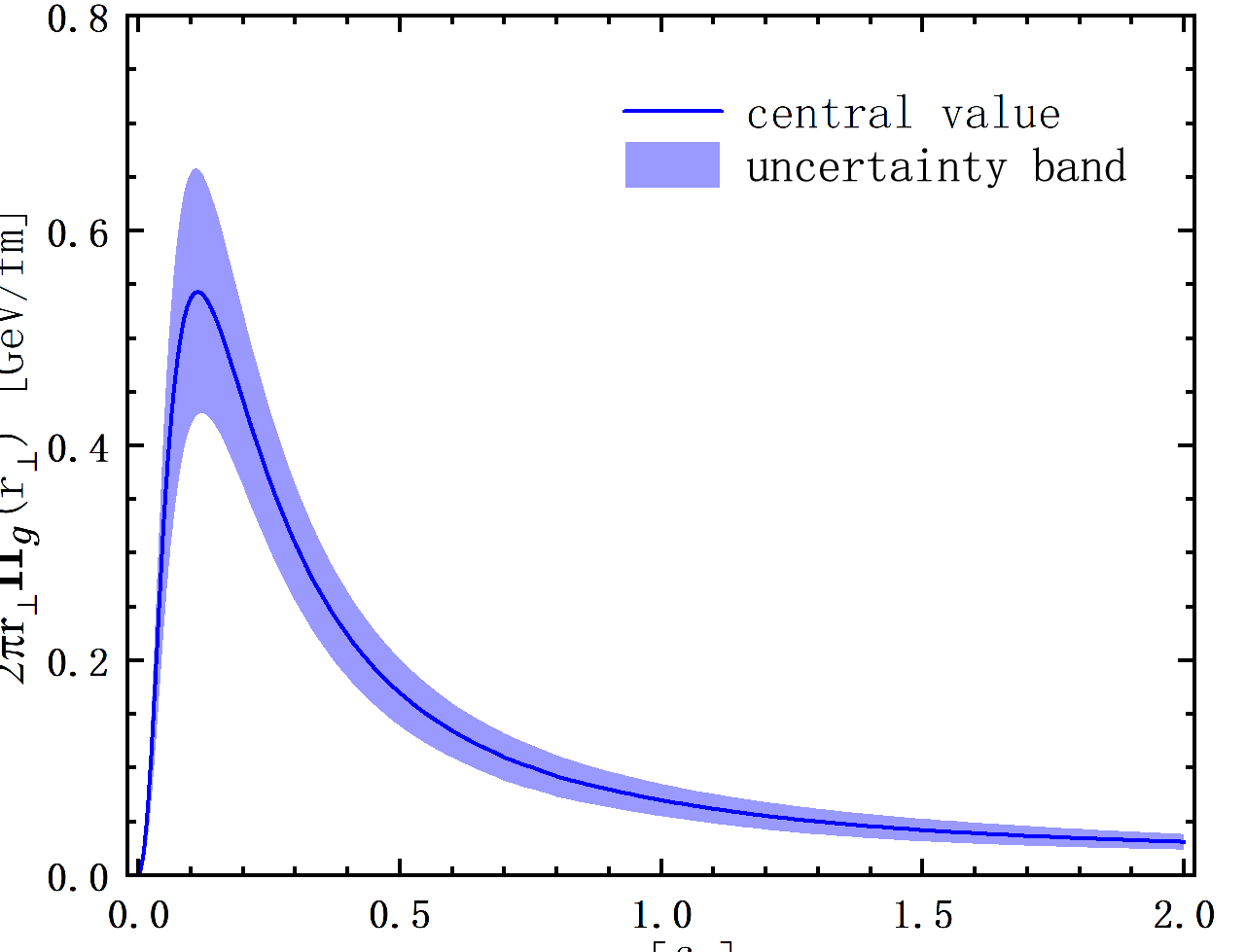}
	\\
	\caption{The 2D gluonic isotropic pressure $\sigma_g(r_\perp)$ (left panel) and pressure anisotropy $\Pi_g(r_\perp)$ (right panel) distributions weighted by $2\pi r_\perp$ as functions of $r_\perp$. The bands represent the parameter uncertainties.}
	\label{fig:sigmapi}
\end{figure}

Figs.~\ref{fig:mu}-\ref{fig:sigmapi} show the 2D gluonic Galilean energy density, radial pressure, tangential pressure, isotropic pressure, and pressure anisotropy, each weighted by $2\pi r_\perp$. 
Qualitatively, these results are consistent with those in the light-front spectator model motivated by soft-wall AdS/QCD~\cite{Sain:2025kup}, but opposite to those in the dressed quark model~\cite{More:2023pcy}.
Quantitatively, our model predictions have a slightly larger magnitude and are more concentrated toward the center of the proton compared to those in Ref.~\cite{Sain:2025kup}. 
Except for $\mu_g(r_\perp)$, similar tendencies also appear in the corresponding quark distributions evaluated using a light-front quark-diquark model~\cite{Chakrabarti:2020kdc} and a BLFQ approach~\cite{Nair:2024fit}, with a wider profile.

Fig.~\ref{fig:mu} displays the gluonic Galilean energy distribution, which is negative near the center, surrounded by a ring of positive energy density that extends outward over a long range, with the negative and positive regions peaking around $r_\perp=0.03\,\text{fm}$ and $r_\perp=0.26\,\text{fm}$, respectively, and the nodal point located at $r_0\approx0.15\,\text{fm}$.
Note that the 2D Galilean energy density of the quark is positive across the entire impact parameter range~\cite{Chakrabarti:2020kdc,Nair:2024fit}.

In Fig.~\ref{fig:sigmart}, the gluonic radial pressure distribution (left panel) remains positive over the entire $r_\perp$ region, indicating a repulsive nature. 
The gluonic tangential pressure distribution (right panel) remains positive (repulsive) at small $r_\perp$ up to $r_\perp\sim0.16\,\text{fm}$, then changes sign and exhibits a negative (attractive) distribution towards the peripheral region, with the repulsive and attractive regions peaking around $r_\perp=0.03\,\text{fm}$ and $r_\perp=0.29\,\text{fm}$, respectively.

Finally, we present the gluonic isotropic pressure distribution and pressure anisotropy in Fig.~\ref{fig:sigmapi}. 
The gluonic isotropic pressure (left panel) is positive for $r_\perp \lesssim 0.24\,\text{fm}$, and is negative at larger $r_\perp$. Since isotropic pressure averages radial and tangential contributions, the peripheral behavior is dominated by the tangential pressure. The pressure anisotropy (right panel), defined as their difference, is positive for all $r_\perp$, meaning radial pressure dominates. This anisotropy peaks near $0.11\,\text{fm}$, while $\Pi_g(r_\perp)$ vanishes at the central position of the proton, consistent with spherical symmetry.

\section{Summary}\label{Sec:5}

In this work, we computed the gluon GFFs and the 2D gluonic mechanical properties of the proton by employing a spectator model, where the proton is described as a two-body composite system consisting of an active gluon and a spectator.
Within the light-front formalism, the GFFs are given by overlaps of the proton wave functions for incoming and outgoing states.
Model parameters are fixed via a global fit to the differential cross section of near-threshold $J/\psi$ photoproduction.
With these fitted parameters and their uncertainties, our gluon GFFs agree well with recent lattice QCD results, and the predicted total cross section is also consistent with experimental data.

We further analyzed the 2D light-front distributions of gluonic mass, angular momentum, pressure, shear force, radial force and tangential force.
We evaluated the mass and pressure distributions at $r_\perp=0\,\text{fm}$, as well as the 2D line tension.
The 2D von Laue stability condition for pressure is qualitatively satisfied, while the one-dimensional von Laue condition for the tangential force and the two-dimensional local stability conditions for shear and radial forces hold quantitatively, supporting the mechanical stability of the proton.
These distributions feature narrower profiles and share qualitative features with gluon distributions from other models and quark distributions extracted from experiment or theory.

Lastly, we examined the 2D gluonic Galilean energy density, radial pressure, tangential pressure, isotropic pressure and pressure anisotropy.
Our predictions show similar shapes but narrower profiles and larger magnitudes relative to those obtained from the soft-wall AdS/QCD motivated light-front spectator model.
In that reference model, the vanishing of the pressure anisotropy at $r_\perp=0\,\text{fm}$ follows directly from spherical symmetry.

\section*{Acknowledgements}
This work is partially supported by the National Natural Science Foundation of China under grant numbers 12447136 and 12150013.

\end{document}